\documentclass[
reprint,
 amsmath,amssymb,
 aps,
]{revtex4-1}

\usepackage{graphicx}
\usepackage{dcolumn}
\usepackage{bm}
\usepackage{mhchem}
\usepackage{color}
\usepackage{dcolumn}
\usepackage[T1]{fontenc}
\usepackage{here}

\begin{document}

\preprint{APS/123-QED}

\title{Fermi surface topology and electronic transport properties of a chiral crystal NbGe$_2$ with strong electron-phonon interaction}

\author{Yoshiki J. Sato$^{1,2}$}
\email{yoshiki_sato@rs.tus.ac.jp}
\author{Ai Nakamura$^{1}$}
\author{Rei Nishinakayama$^{2}$}
\author{Ryuji Okazaki$^{2}$}
\author{Hisatomo Harima$^{3}$}
\author{Dai Aoki$^{1}$}
\affiliation{%
 $^{1}$Institute for Materials Research, Tohoku University, Oarai, Ibaraki 311-1313, Japan\\
 $^{2}$Department of physics, Faculty of Science and Technology, Tokyo University of Science, Noda, Chiba 278-8510, Japan\\
 $^{3}$Graduate School of Science, Kobe University, Kobe 657-8501, Japan
}%

\date{\today}

\begin{abstract}
We report the electronic structures and transport properties of a chiral crystal NbGe$_2$, which is a candidate for a coupled electron-phonon liquid. The electrical resistivity and thermoelectric power of NbGe$_2$ exhibit clear differences compared to those of NbSi$_2$ even though both niobium ditetrelides are isostructural and isoelectronic. We discuss the intriguing transport properties of NbGe$_2$ based on a van Hove-type singularity in the density of states. The analysis of de Haas-van Alphen oscillations measured by the field modulation and magnetic torque methods reveals the detailed shape of the Fermi surface of NbGe$_2$ by comparison with the results of energy band structure calculations using a local density approximation. The electron and hole Fermi surfaces of NbGe$_2$ split into two because of the anti-symmetric spin-orbit interaction. The temperature dependence of quantum oscillations indicates that the effective mass is isotropically enhanced in NbGe$_2$ due to strong electron-phonon interaction.
\end{abstract}

\maketitle
\section{INTRODUCTION}
Chiral crystals have attracted great attention as a platform for a wide variety of interesting physical phenomena including magnetochiral dichroism \cite{GLJAR97}, chiral magnetism \cite{SM09,YT12,YJS22,YJS23}, non-reciprocal electronic transport \cite{GLJAR01}, and topologically nontrivial band structures \cite{SST17,GC18}. In particular, a topological electronic structure, which is called Kramers-Weyl fermion, is well defined at time-reversal invariant momenta in non-magnetic chiral crystals \cite{GC18}. The unique electronic structures in non-magnetic chiral crystals, such as B20-type silicides \cite{GC17,PT17,TZ18} and elemental tellurium\cite{TF17,MS20,GG20}, have been intensively studied. When the topological degeneracies exist in the vicinity of the Fermi level, exotic electronic transport and quantized responses are expected. 

\begin{figure}[htbp]
\centering
\includegraphics[width=0.75\linewidth]{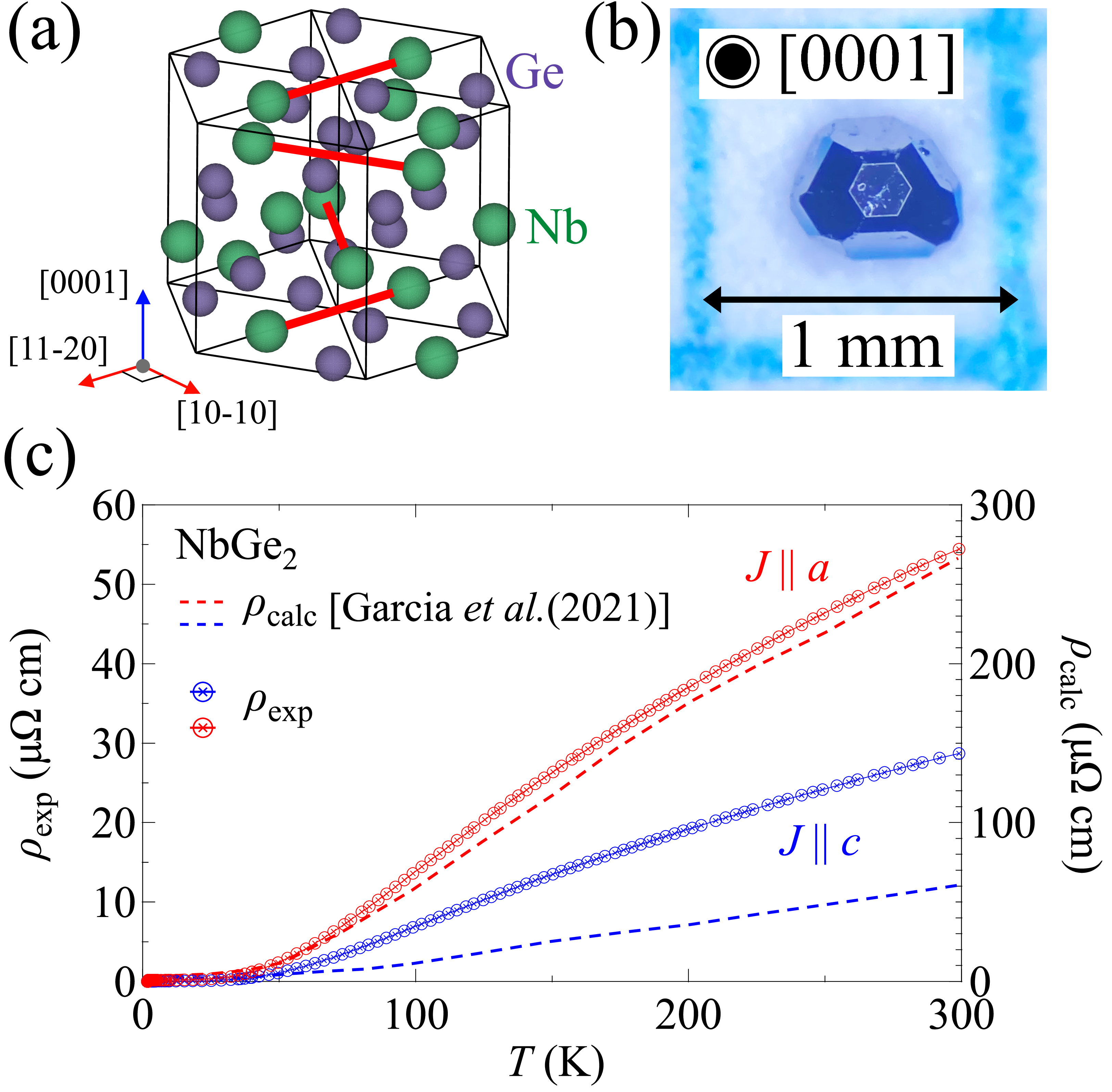}
\caption{\label{fig1}(a) Crystal structure of NbGe$_2$. (b) Single crystal of NbGe$_2$. (c) Temperature dependence of resistivity $\rho_{\rm exp}$ of NbGe$_2$ for $J$ $||$ $[10\bar{1}0]$ and $[0001]$ (left axis). The dashed lines show calculated temperature dependence of resistivity $\rho_{\rm calc}$ of NbGe$_2$ (right axis) from Ref.\cite{CACG21}.}
\end{figure}

In this study, we focus on a transition metal digermanide NbGe$_2$ with a chiral crystal structure [space group $P6_222$ (No. 180, $D^4_6$) or $P6_422$ (No. 181, $D^5_6$)]. NbGe$_2$ belongs to C40-type transition metal ditetrelides $TX_2$ ($T$ = V, Cr, Nb, Ta and $X$ = Si, Ge). The C40-type hexagonal crystal structure of NbGe$_2$ is shown in Fig.~\ref{fig1}(a). NbGe$_2$ is a superconductor with $T_{\rm c}$ = 2 K \cite{JPR78}, and the prediction that NbGe$_2$ belongs to the Kramers-Weyl semimetal\cite{GC18} promoted detailed studies of its superconducting properties including the superconducting phase diagram, full-gap structure, and crossover from type-${\rm I}$ to type-${\rm I\hspace{-.01em}I}$/1 superconductivity \cite{BL20,EE20,DZ21}. Several previous studies have reported energy band structures obtained by first-principles calculations taking into account the spin-orbit coupling (SOC), and doubly degenerate Kramers-Weyl points appear near the Fermi energy level at M and L points \cite{GC18,EE20,BL20,CACG21}.

Recent theoretical and experimental studies of NbGe$_2$ have shown that strong electron-phonon interaction exists, and NbGe$_2$ has been deemed to be an ideal candidate to observe phonon-mediated hydrodynamic flow \cite{CACG21,HYY21}. When the momentum-conserving scattering process is dominant compared to the momentum-relaxing one, the hydrodynamic electron flow \cite{AL20}, as seen in a metallic delafossite PdCoO$_2$ \cite{PJWM16} and Weyl semimetals \cite{JG18,AJ18,UV21}, can be observed. A previous theoretical study using the first-principles calculation and the Boltzmann transport equation has pointed out that the momentum-conserving phonon-mediated electron-electron scattering lifetime is dominant, and an anisotropic scattering time leads to an anisotropic electronic transport in NbGe$_2$ \cite{CACG21}. A subsequent experimental study using quantum oscillations and Raman scattering experiments has demonstrated the strong electron-phonon interaction and strong suppression of momentum-relaxing phonon-phonon processes in NbGe$_2$ \cite{HYY21}. However, little is known about the topology of bulk Fermi surfaces and anisotropy of the effective masses due to the anisotropic electron-phonon scattering time in NbGe$_2$.

In this paper, we report on the topology of the Fermi surface of NbGe$_2$ and the transport properties closely related to the characteristic electronic structure. NbGe$_2$ and NbSi$_2$ exhibit a notable difference in the electronic transport and thermoelectric properties due to the characteristic electronic structure of  NbGe$_2$ although both compounds are isostructural and isoelectronic. Then, we report the de Haas-van Alphen (dHvA) experiments of NbGe$_2$. In spite of rather complicated Fermi surfaces, a limited number of dHvA branches has been observed experimentally \cite{EE20,HYY21}. To fully reveal the Fermi surface topology of NbGe$_2$, we have performed two different techniques: magnetic field modulation and magnetic torque methods. By comparing experimental results with energy band structure calculation, the detailed shape of the Fermi surface is clarified. The enhancement of effective masses in NbGe$_2$ has been demonstrated by the analysis of temperature dependence of quantum oscillations. The effective mass enhancement and electronic transport properties indicate the presence of strong electron-phonon interaction in NbGe$_2$.

\section{EXPERIMENTAL DETAILS}
The polycrystalline samples of NbGe$_2$ and NbSi$_2$ were synthesized via arc melting using an arc furnace under an argon atmosphere. The starting materials were Nb (99.9 \%), Ge (99.999 \%), and Si (99.9999 \%) with a stoichiometric ratio, namely Nb:Ge = 1:2 and Nb:Si = 1:2. Single crystals of NbGe$_2$ were grown using the chemical vapor transport (CVT) technique. The NbGe$_2$ polycrystal was crushed into small pieces and sealed in a quartz tube with 80 mg iodine as a transport agent. The sealed quartz tube was placed in a furnace with a temperature gradient of 950 $^\circ$C/850 $^\circ$C for two weeks. The NbGe$_2$ single crystals were grown at the high-temperature side. The typical size of the grown crystal is up to the dimensions of 0.6 mm $\times$ 0.6 mm $\times$ 0.6 mm, as shown in Fig.$~$\ref{fig1}(b).
Single crystals of NbSi$_2$ were grown using the Czochralski method in a tetra-arc furnace.
The crystal structure of NbGe$_2$ single crystals grown by CVT was confirmed using a single-crystal x-ray diffractometer (Rigaku XtaLAB mini ${\rm I\hspace{-.01em}I}$) with Mo $K\alpha$ radiation ($\lambda$ = 0.71073 $\mathrm{\mathring{A}}$). The crystal structure was solved with SHELXT and then refined with SHELXL. The refined crystallographic parameters and atomic positions of NbGe$_2$ are summarized in Tables \ref{t1} and \ref{t2}. The single crystals were oriented using a Laue camera (Photonic Science Laue x-ray CCD camera).
\begin{table}[t]
\caption{\label{t1} Crystallographic and structural refinement data of NbGe$_2$ obtained from single-crystal XRD.}
\begin{ruledtabular}
\begin{tabular}{lc}
Empirical formula & NbGe$_2$\\
Formula weight & 238.09\\
Crystal system &hexagonal\\
Space group & $P6_422$ (\#181)\\
$a$ ($\mathrm{\mathring{A}}$) & 4.9727(7) \\ 
$c$ ($\mathrm{\mathring{A}}$) & 6.7872(10) \\ 
Volume ($\mathrm{\mathring{A}}^3$) & 145.35(4) \\ 
Formula units per cell (Z) & 3 \\
Number of measured reflections (total) & 1177\\
Number of measured reflections (unique) & 191\\
Cut off angle (2$\theta_{\rm max}$) & 65.9$^\circ$\\
$R1$ ($I$ $>$ 2.00$\sigma$($I$)) & 0.0205\\
$R$ (All reflections) & 0.0291\\
$wR2$ (All reflections) & 0.0470\\
Goodness of fit & 1.165\\
Flack parameter & 0.01(3)\\
Max Shift/Error in Final Cycle & 0.000\\
\end{tabular}
\end{ruledtabular}
\end{table}
\begin{table}[b]
\caption{\label{t2}Atomic positions and displacement parameters of NbGe$_2$.}
\begin{ruledtabular}
\begin{tabular}{l c c c c c}
\multicolumn{6}{c}{NbGe$_2$}\\ \hline
Atom & Site & {\it x} & {\it y} & {\it z} & $B_{eq}$\\ \hline
Nb & 3$c$ & 1/2 & 0 & 0 & 0.37(3) \\
Ge & 6$i$ & 0.16444(8) & 0.32888(8) & 0 & 0.31(2) \\
\end{tabular}
\end{ruledtabular}
\end{table}

The electrical resistivity was measured with a Quantum Design physical property measurement system (QD PPMS). The thermopower was measured by a steady-state technique using a manganin-constantan thermocouple\cite{TY22} in a GM refrigerator. The temperature gradient ($\sim 0.5$ K/mm) was applied along the in-plane direction using a resistive heater.
The dHvA effects were detected using the magnetic field modulation and magnetic torque methods. The magnetic field modulation measurements were performed using a top-loading dilution refrigerator in magnetic fields up to 14.7 T at low temperatures down to 30 mK. The magnetic torque was measured using a membrane-type surface stress sensor in magnetic fields up to 9 T at low temperatures down to 1.7 K.

The band structure calculations were performed using the KANSAI code based on the full-potential linearized augmented plane wave (FLAPW) method within the local density approximation (LDA). In the band calculations, the scalar relativistic effect is considered for all electrons. The spin-orbit coupling is included by means of a second variational procedure for all valence electrons. The lattice parameters and atomic positions used for the calculations are listed in Tables \ref{t1} and \ref{t2}.

\begin{figure*}[htbp]
\centering
\includegraphics[width=\linewidth]{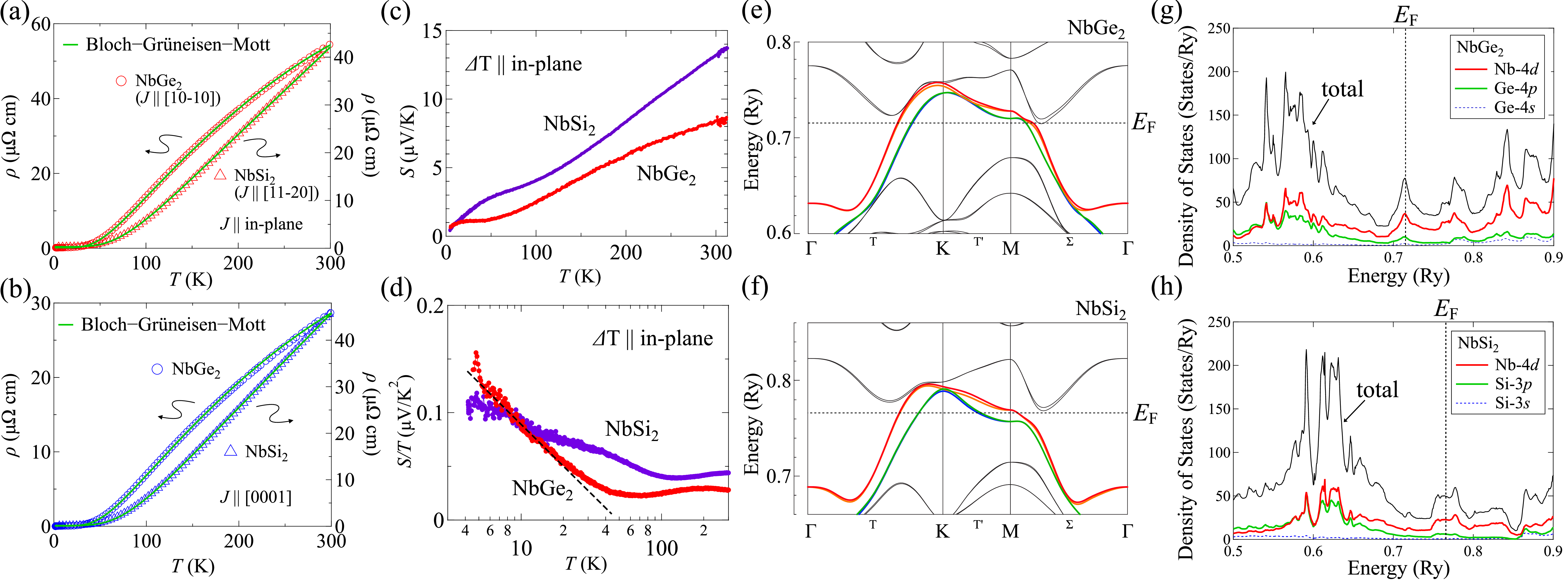}
\caption{\label{fig2} Comparison of $\rho(T)$ of NbGe$_2$ and NbSi$_2$ for (a) in-plane and (b) out-of-plane current directions. The solid lines are fitting results based on Eq.~(\ref{BGM}). (c) Temperature dependence of the Seebeck coefficient $S$ of NbGe$_2$ and NbSi$_2$. The temperature gradient is along the in-plane direction. (d) $S$/$T$ of NbGe$_2$ and NbSi$_2$ on a logarithmic $T$ scale. Electronic band structure of (e) NbGe$_2$ and (f) NbSi$_2$ along the $\Gamma$-K-M-$\Gamma$ cut. Calculated total and partial density of states for (g) NbGe$_2$ and (h) NbSi$_2$.}
\end{figure*}

\section{RESULTS and DISCUSSION}
We have measured the temperature dependence of electrical resistivity $\rho(T)$ of a NbGe$_2$ single crystal for the electrical current $J$ $||$ $[10\bar{1}0]$ (in-plane) and $||$ $[0001]$ (out-of-plane), as shown in Fig.$~$\ref{fig1}(c). At room temperature, $\rho_{\rm 300\,K}$ = 54 ${\rm \mu\Omega\, cm}$ for $J$ $||$ $[10\bar{1}0]$ and 29 ${\rm \mu\Omega\, cm}$ for $J$ $||$ $[0001]$. The anisotropy of $\rho$ is approximately 2 between the in-plane and out-of-plane current directions at 300 K. The residual resistivity ratio (RRR $\equiv$ $\rho_{\rm 300\,K}$/$\rho_0$) is 320 ($\rho_0$ = 0.17 ${\rm \mu\Omega\, cm}$) for $J$ $||$ $[10\bar{1}0]$ and 640 ($\rho_0$ = 0.05 ${\rm \mu\Omega\, cm}$) for $J$ $||$ $[0001]$. The small values of $\rho_0$ and the large values of RRR indicate the high quality of the single crystals of NbGe$_2$ in this study.

The calculated electrical resistivity obtained using the Boltzmann transport equation within the relaxation time approximation \cite{CACG21} is also shown in Fig.$~$\ref{fig1}(c). Except for a difference in the absolute values, $\rho_{\rm exp}(T)$ and $\rho_{\rm calc}(T)$ show similar behavior including the convex upward curvature. As shown in Fig.$~$\ref{fig1}(c), previous theoretical calculations have predicted much larger and more anisotropic electrical resistivity of NbGe$_2$ at room temperature because of the strong electron-phonon scattering. The discrepancy between $\rho_{\rm exp}(T)$ and $\rho_{\rm calc}(T)$ may be due to the dominance of momentum-conserving scattering processes in NbGe$_2$ at high temperatures, as discussed in the previous experimental work \cite{HYY21}.

Here, we show a notable difference in the temperature dependence of electrical resistivity of NbGe$_2$ and NbSi$_2$, both of which are isostructural and isoelectronic. Figures \ref{fig2}(a) and \ref{fig2}(b) show a comparison of $\rho(T)$ between NbGe$_2$ and NbSi$_2$ for in-plane and out-of-plane current directions, respectively. NbGe$_2$ exhibits distinct anisotropy in electrical resistivity, while NbSi$_2$ shows isotropic electrical resistivity. The electrical resistivity of NbSi$_2$ and NbGe$_2$ are well fitted by the Bloch--Gr\"{u}neisen--Mott formula \cite{NFM36,AHW38,NFM58} as:
\begin{equation}
\rho(T) = \rho_0 + A\left(\frac{T}{\theta_{\rm D}}\right)^5\int_0^{\theta_{\rm D}/T}\frac{x^5}{(e^x-1)(1-e^{-x})}dx - KT^3,
\label{BGM}
\end{equation}
where the second term and the third term describes the electron-phonon intraband scattering and the electron-phonon interband ($s$-$d$) scattering, respectively. We obtained the Debye temperature $\theta_{\rm D}$ of NbSi$_2$ as $\theta_{\rm D} = 460$ K ($\theta_{\rm D} = 485$ K) for $J$ $||$ in-plane (out-of-plane) and $\theta_{\rm D}$ of NbGe$_2$ as $\theta_{\rm D} = 315$ K ($\theta_{\rm D} = 330$ K) for $J$ $||$ in-plane (out-of-plane). The Debye temperature is isotropic in both NbSi$_2$ and NbGe$_2$.
The obtained coefficients for intraband ($A$) and interband ($K$) electron-phonon scattering of NbSi$_2$ are as follows: $A = 280$ ${\rm \mu\Omega cm}$ ($A = 330$ ${\rm \mu\Omega cm}$ ), and $K = -6 \times10^{-8}$ ${\rm \mu\Omega cm/K^3}$ ($K = -5 \times 10^{-8}$ ${\rm \mu\Omega cm}$) for in-plane (out-of-plane) current direction. Similarly, the coefficients for NbGe$_2$ are as follows: $A = 290$ ${\rm \mu\Omega cm}$ ($A = 160$ ${\rm \mu\Omega cm}$ ), and $K = 4 \times10^{-7}$ ${\rm \mu\Omega cm/K^3}$ ($K = 2 \times 10^{-7}$ ${\rm \mu\Omega cm}$) for in-plane (out-of-plane) current direction. The coefficients obtained from fitting the Eq.(\ref{BGM}) are summarized in Table~\ref{t2p5}. The intraband electron-phonon scattering is comparable and rather isotropic in both NbSi$_2$ and NbGe$_2$. Meanwhile, the coefficients for the interband electron-phonon scattering in NbGe$_2$ differ by an order of magnitude from those in NbSi$_2$, indicating the interband $s$-$d$ scattering process is dominant in NbGe$_2$. This significant interband electron-phonon scattering is the origin of the pronounced convex temperature dependence in the resistivity of NbGe$_2$. In addition, the sign of the coefficient $K$ is reversed between NbSi$_2$ and NbGe$_2$, reflecting the difference in $d$-band electronic structure \cite{NFM58}.
\begin{table*}[htbp]
\caption{\label{t2p5} Coefficients obtained from fitting the Bloch--Gr\"{u}neisen--Mott formula for NbSi$_2$ and NbGe$_2$.}
\begin{ruledtabular}
\begin{tabular}{l c c c c c}
 &  & $\rho_0$ (${\rm \mu\Omega\ cm}$) & $\theta_{\rm D}$ (K) & $A$ (${\rm \mu\Omega\ cm}$) & $K$ (${\rm \mu\Omega\ cm/K^3}$)\\ \hline
NbSi$_2$ & in-plane & 0.322 & 460 & 280 & $-6\times10^{-8}$ \\
 & out-of-plane & 0.167 & 485 & 330 & $-5\times10^{-8}$ \\
NbGe$_2$ & in-plane & 0.165 & 315 & 290 & $4\times10^{-7}$ \\
 & out-of-plane & 0.050 & 330 & 160 & $2\times10^{-7}$ \\
\end{tabular}
\end{ruledtabular}
\end{table*}

In addition, intriguing differences in thermoelectric properties were observed between NbGe$_2$ and NbSi$_2$. Figure$~$\ref{fig2}(c) shows the temperature dependence of Seebeck coefficient $S$ of NbGe$_2$ and NbSi$_2$ for $\Delta T$ $||$ in-plane. $S(T)$ varies linearly with temperature above 150 K. Figure$~$\ref{fig2}(d) shows a comparison of $S(T)/T$ between NbGe$_2$ and NbSi$_2$. $S(T)/T$ of NbGe$_2$ showed a metallic behavior ($S(T)/T =$ cosnt.) at high temperatures, while $S(T)/T$ is enhanced with a logarithmic temperature dependence ($S(T)/T \propto -{\rm ln} T$) at low temperatures. In a high temperature region, $S(T)$ of NbGe$_2$ is smaller than that of NbSi$_2$, but exceeds the $S(T)$ of NbSi$_2$ below approximately 9 K.

To discuss the characteristic electronic transport properties of NbGe$_2$, we performed LDA calculations for energy band structures of NbGe$_2$ and NbSi$_2$. Figures$~$\ref{fig2}(e) and \ref{fig2}(f) show the energy band structures obtained by LDA calculations of NbGe$_2$ and NbSi$_2$. The parity violation, namely the lack of inversion symmetry in the chiral crystal, leads to the spin-splitting of energy bands. However, two-fold degenerate points, namely the Kramers-Weyl points, exist at high-symmetry points of the Brillouin zone, such as $\Gamma$ and M points. The calculated energy band structure is essentially consistent with the energy band structure reported in previous studies \cite{BL20,EE20,HYY21}.

The notable difference in the electrical resistivity and the Seebeck coefficient between NbGe$_2$ and NbSi$_2$ may be attributed to the difference in the energy dependence of the density of states $D(\varepsilon)$ near the Fermi energy $E_{\rm F}$. The calculated density of states is shown in Figs.$~$\ref{fig2}(g) and \ref{fig2}(h). A van Hove-type distinct peak structure appears in $D(\varepsilon)$ of NbGe$_2$ near $E_{\rm F}$. In addition, a contribution from Nb-4$d$ electrons is significant in the $D(E_{\rm F})$ of NbGe$_2$. Since the density of states in the $d$ band is much larger than in the $s$ ($p$) state, the final density of states is higher when the $s$ ($p$) electrons scatter to the $d$ state. The dominant interband electron-phonon scattering process in $\rho(T)$ of NbGe$_2$ is most likely due to the large contribution of Nb-4$d$ electrons in the vicinity of $E_{\rm F}$. In the archetypal example, the electrical resistivity of Pd exhibits a convex temperature dependence due to the peak structure in the DOS resulting from the contribution of the Pd-4$d$ electrons \cite{NFM58,FMM70,FYF74}.

The enhancement of low-temperature $S(T)/T$ of NbGe$_2$ is also attributed to $D(\varepsilon)$. The Mott formula that describes the Seebeck coefficient in metals is expressed as \cite{MC69}:
\begin{equation}
S = \frac{\pi^2}{3}\frac{k_{\rm B}}{q}(k_{\rm B}T)\left.\frac{\partial{\rm ln}(D)}{\partial \varepsilon}\right|_{\varepsilon = E_{\rm F}},
\label{Mott}
\end{equation}
where $k_{\rm B}$ and $q$ are the Boltzmann constant and the carrier charge, respectively. The enhancement of low-temperature $S/T$ of NbGe$_2$ is probably due to the large value of $\partial{\rm ln}(D)/\partial \varepsilon |_{\varepsilon = E_{\rm F}}$, while the flat $D(\varepsilon)$ near $ E_{\rm F}$ results in a small low-temperature $S/T$ of NbSi$_2$. Note that a peak in $S(T)/T$ due to the phonon-drag effect has been observed at around 20 K in the previous study \cite{HYY21}, while the phonon-drag effect is absent in $S(T)/T$ of the single crystal used in this study. The difference in $S(T)/T$ may be attributed to the surface scattering effect depending on the sample size \cite{HT16}.

\begin{figure}[b]
\centering
\includegraphics[width=\linewidth]{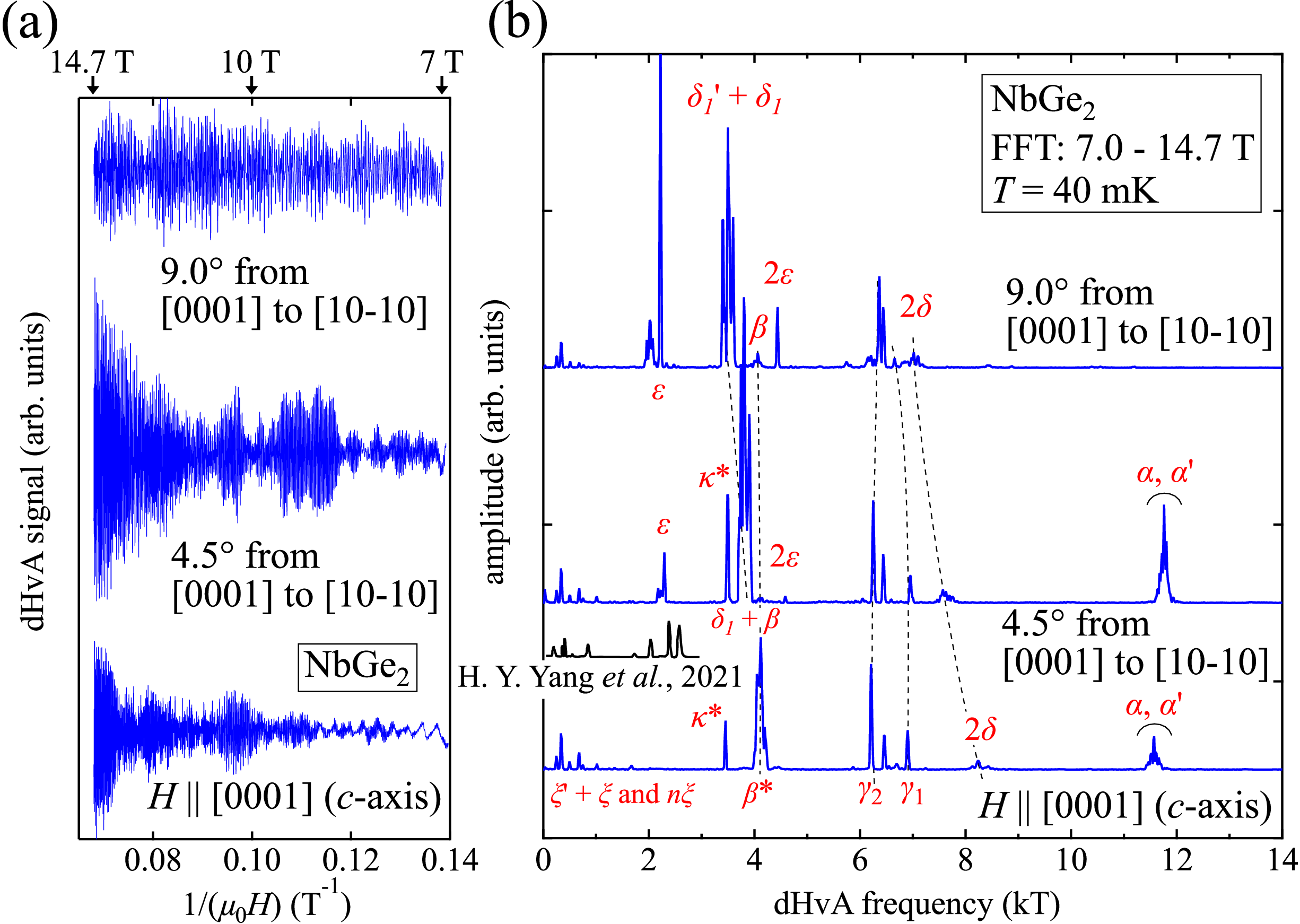}
\caption{\label{fig3} Typical dHvA oscillations (a) and corresponding FFT spectra (b) of NbGe$_2$ at $T$ = 40 mK. The dHvA oscillations were detected using the field modulation technique, and the magnetic field range of FFT is 7.0 to 14.7 T. The FFT spectrum reported in a previous work (Ref.\cite{HYY21}) is also shown in panel (b).}
\end{figure}

The dHvA measurements of NbGe$_2$ single crystals were performed to reveal the topology of Fermi surfaces of NbGe$_2$. The dHvA frequency $F$ is proportional to the extremal cross-sectional area of the Fermi surface $S_{\rm F}$, that is, $F$ = $(\hbar c/2\pi e)S_{\rm F}$. Figure \ref{fig3}(a) shows the typical dHvA oscillations of NbGe$_2$ detected using the field modulation technique, and the corresponding fast Fourier transform (FFT) spectra are shown in Fig.$~$\ref{fig3}(b).
Several dHvA frequencies appeared in the FFT spectra. The FFT peaks $\zeta$, $\zeta'$, $\kappa^\ast$, $\varepsilon$, $\delta_1$, $\delta_1'$, $\beta$, $\beta^\ast$, $\gamma_1$, $\gamma_2$, $\alpha$, and $\alpha'$ are fundamental signals in Fig.~\ref{fig3}(b). The harmonic signals $n\zeta$ ($n$ = 1, 2, and 3), 2$\varepsilon$, and 2$\delta$ were also observed. The dHvA branches $\zeta$, $\zeta'$, and $\varepsilon$ have been observed in magnetic torque measurements in previous studies \cite{EE20,HYY21}. The critical field of NbGe$_2$ is $H_{\rm c2}$($T$ = 0) $\sim$ 0.04 T \cite{BL20,EE20}, and dHvA measurements in this study were performed in the normal state.

\begin{figure}[b]
\centering
\includegraphics[width=\linewidth]{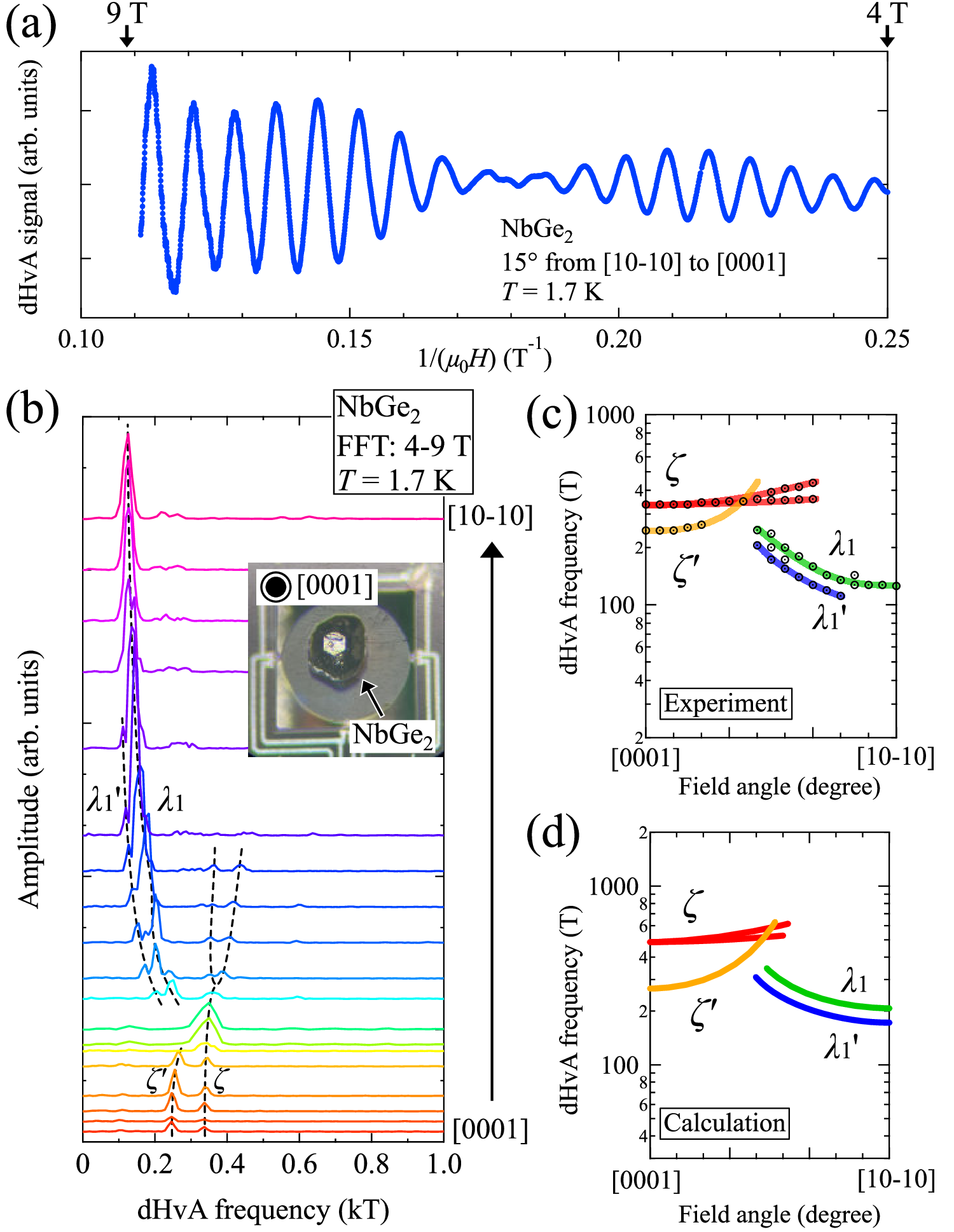}
\caption{\label{fig4} The dHvA oscillations (a) and corresponding FFT spectra (b) obtained from magnetic torque measurements. The field-angle is from $[0001]$ to $[10\bar{1}0]$ in steps of 5$^\circ$. (c) Angular dependence of experimental $\zeta-\zeta'$ and $\lambda_1-\lambda_1'$ branches. (d) Angular dependence of calculated dHvA frequencies of NbGe$_2$.}
\end{figure}

In the field modulation technique, the Bessel function of the first kind in the detected dHvA signal makes it difficult to observe small Fermi surfaces. The magnetic torque measurement is a suitable method to detect the small Fermi surface since the dHvA oscillations in the magnetic torque are not affected by the Bessel function. Therefore, we have performed dHvA experiments of NbGe$_2$ using the magnetic torque method. Figure$~$\ref{fig4}(a) shows the FFT spectra of the dHvA oscillations for the magnetic field directions from $[0001]$ to [10$\bar{1}$0].
\begin{figure*}[t]
\centering
\includegraphics[width=\linewidth]{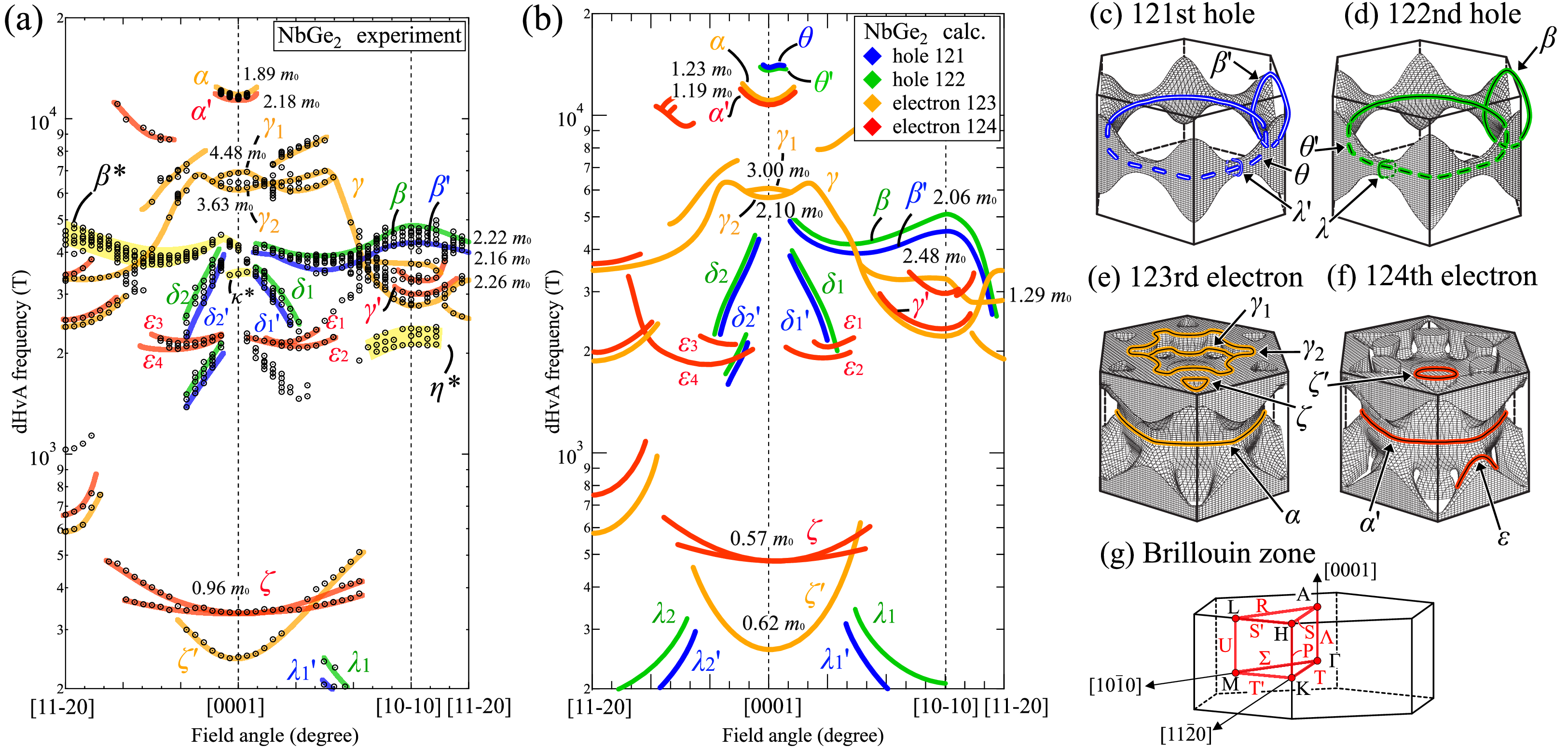}
\caption{\label{fig5}Comparison of (a) experimental dHvA frequencies of NbGe$_2$ with (b) calculated dHvA frequencies obtained by the LDA calculations. The solid lines in panel (a) are guides to the eyes. The cyclotron effective mass and the band masses are denoted for the principal directions. (c)--(f) Calculated Fermi surfaces of NbGe$_2$. (c) and (d): Spin-split hole Fermi surfaces. (e) and (f): Spin-split electron Fermi surfaces. Solid and dashed lines on the Fermi surfaces indicate the extremal cross-sections. (g) Brillouin zone of the hexagonal lattice. The symmetry points and axes are denoted by their common names.}
\end{figure*}
Figure \ref{fig4}(b) shows the field angle dependence of the experimental dHvA frequencies of NbGe$_2$. In addition to the $\zeta$ and $\zeta'$ branches observed in the field modulation method, dHvA branches $\lambda_1$ and $\lambda_1'$ have been observed in the magnetic torque measurement. Note that the labels of dHvA branches are different from that in a previous study: $\lambda$ and $\zeta$ correspond to $\alpha$ and $\beta$ in Ref.\cite{EE20}, respectively. As discussed below, the existence of $\lambda_1$ and $\lambda_1'$ branches and the angular dependence provide important insights into the topology of the Fermi surface.

\begin{figure}[t]
\centering
\includegraphics[width=\linewidth]{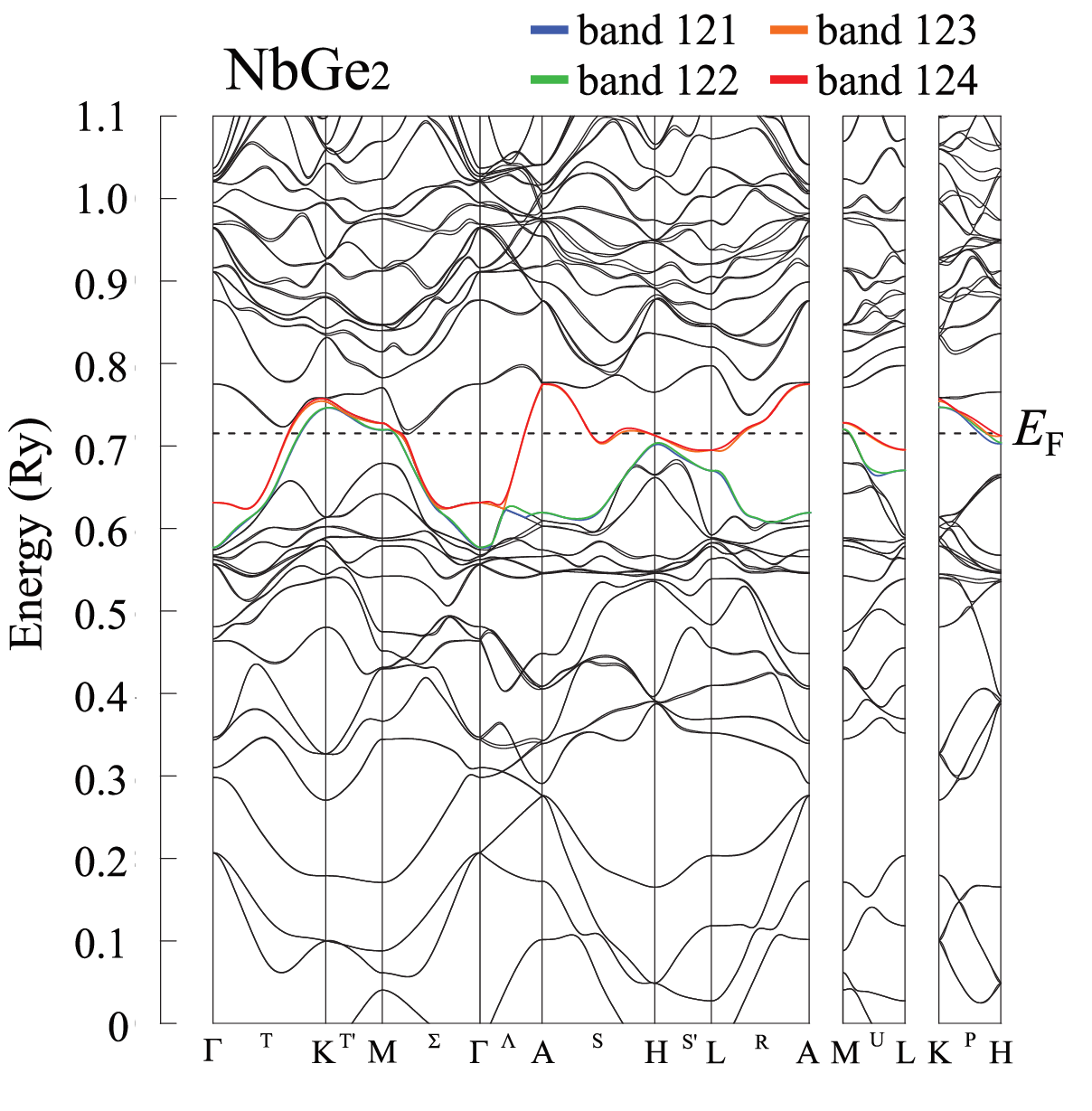}
\caption{\label{fig6}Calculated band structure of NbGe$_2$. The dashed line indicates the Fermi level.}
\end{figure}

Here, we compare the experiment and calculated angular dependence of dHvA frequencies. Figure \ref{fig5}(a) shows the experimental angular dependence of the dHvA frequencies obtained using the field modulation and magnetic torque methods. To interpret the observed dHvA frequencies, we calculated the Fermi surface using the result of LDA calculations. Figure$~$\ref{fig6} shows the calculated energy bands of NbGe$_2$.
The calculated Fermi surfaces of NbGe$_2$ are shown in Figs.$~$\ref{fig5}(c)--\ref{fig5}(f). The Fermi surfaces of NbGe$_2$ consist of spin-split electron and spin-split hole Fermi surfaces. Then, we constructed the angular dependence of the dHvA frequency from the calculated Fermi surface, as shown in Fig.\ref{fig5}(b). The angular dependence of the experimental dHvA frequencies [Fig.$~$\ref{fig5}(a)] can be interpreted based on the calculated result [Fig.$~$\ref{fig5}(b)] although some dHvA branches, such as $\beta^\ast$, $\eta^\ast$, and $\kappa^\ast$, appeared only in the experimental result. 

$\alpha$, $\gamma$, $\varepsilon$, and $\zeta$ branches correspond to the pairs of spin-split electron Fermi surfaces [Figs.$~$\ref{fig5}(e) and \ref{fig5}(f)]. The field angle dependence of these dHvA branches corresponding to the electron Fermi surfaces is in good agreement with the calculations. On the other hand, dHvA branches corresponding to the spin-split hole Fermi surfaces, such as $\theta$, $\beta$, $\delta$, and $\lambda$, show discrepancies between the results of experiments and LDA calculations. The $\theta$ and $\theta'$ branches corresponding to the inner paths of the hole Fermi surfaces were not observed in the experiment. In the LDA calculation, we have obtained the relatively large curvature factors, namely the term of ($\partial^2S_{\rm F}/\partial k^2$)$^{-1/2}$ in the dHvA amplitude, for $\theta$ and $\theta'$ branches. The large curvature factors for the inner orbits indicate that these branches are difficult to observe experimentally. Instead of the $\theta$ and $\theta'$ branches, the $\lambda$ branches corresponding to the small neck near the $M$ point provide evidence for the existence of the connected hole Fermi surfaces. As mentioned above, the small dHvA branches $\lambda_1$ and $\lambda_1'$ have been detected by using the magnetic torque method. Figure$~$\ref{fig4}(d) shows the angular dependence of the calculated dHvA frequencies of NbGe$_2$ in the region of low dHvA frequencies. The experimental result [Fig.$~$\ref{fig4}(c)] and calculated angular dependence [Fig.$~$\ref{fig4}(d)] show a good agreement, indicating the existence of the spin-split hole surfaces with the shapes shown in Figs.$~$\ref{fig5}(c) and \ref{fig5}(d).

Furthermore, the experimentally observed $\beta$ branch shows a different angular dependence from the calculated one. As shown in Fig.$~$\ref{fig5}(a), the $\beta$ and $\beta'$ branches have been observed for all field-angle in the basal plane, namely $[10\bar{1}0]$ to $[11\bar{2}0]$, whereas the calculated $\beta$ and $\beta'$ branches were absent for $H$ $||$ $[11\bar{2}0]$ [Fig.$~$\ref{fig5}(b)] because of the small neck of hole Fermi surfaces. Moreover, the $\beta^\ast$ branch has been observed for the field directions from $[0001]$ to $[11\bar{2}0]$ in the field modulation measurements although the $\beta^\ast$ branch does not appear in the LDA calculation. This difference between experiments and calculations is likely due to the magnetic breakdown on the small neck orbit at around the M point. We note that many dHvA frequencies have been observed between $\beta$ and $\beta'$ branches. The many dHvA frequencies can be interpreted as being caused by the magnetic breakdown effect in the two spin-split Fermi surfaces.

\begin{figure}[t]
\centering
\includegraphics[width=\linewidth]{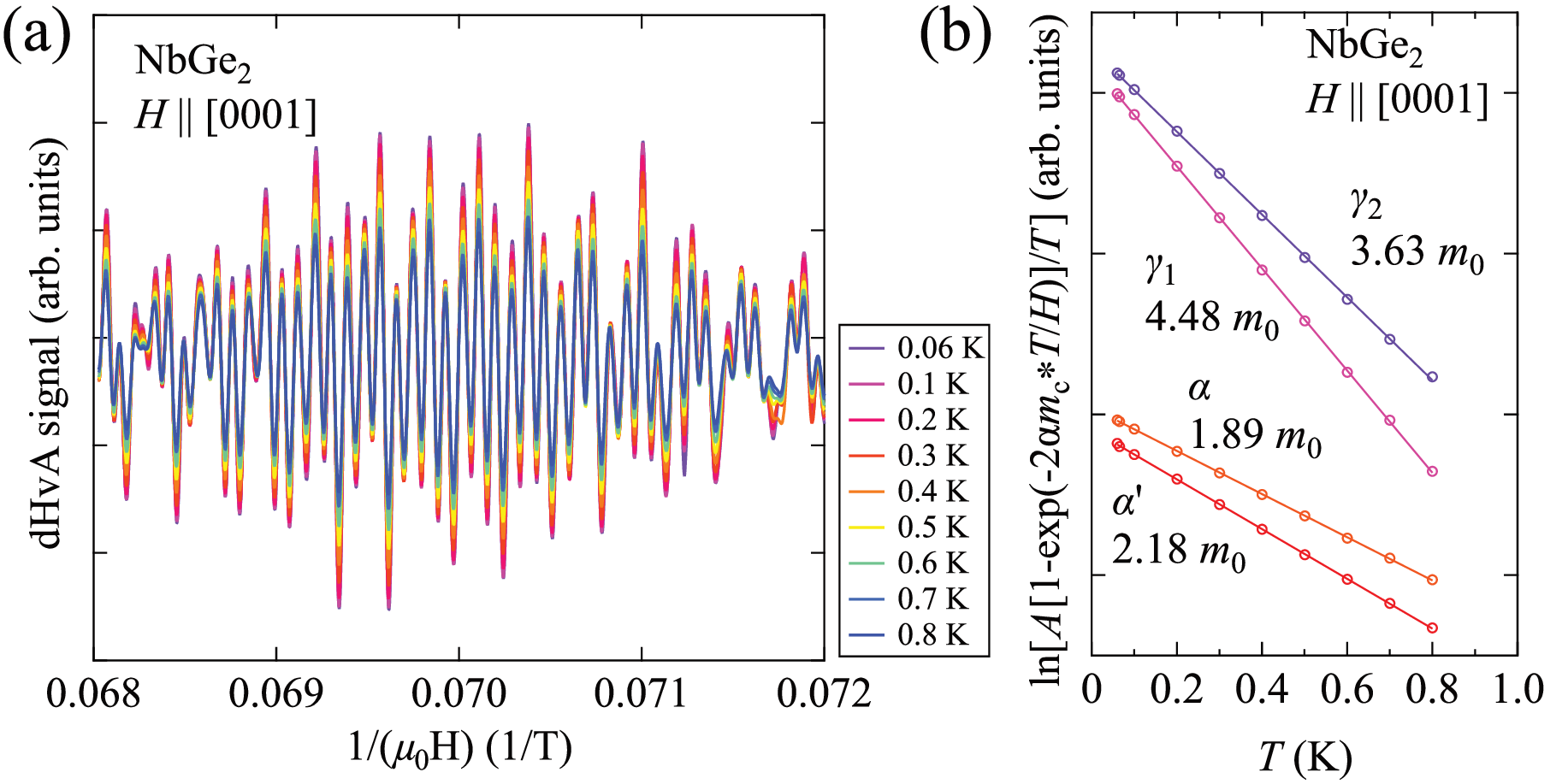}
\caption{\label{fig7}(a) The dHvA oscillations measured at several constant temperatures for $H$ $||$ $[0001]$. (b) Mass plot of $\alpha$, $\alpha'$, $\gamma_1$, and $\gamma _2$ branches.}
\end{figure}

To obtain the cyclotron effective mass, we have measured dHvA oscillations at several constant temperatures in the range from 0.06 to 0.8 K for three field directions: $[0001]$, $[10\bar{1}0]$, and $[11\bar{2}0]$. Figure \ref{fig7}(a) shows the dHvA oscillations for $H$ $||$ $[0001]$, demonstrating that the dHvA amplitude decays as the temperature increases.
Based on the Lifshitz-Kosevich (LK) formula, the temperature dependence of dHvA amplitude can be written as follows \cite{DS84}:
\begin{equation}
{\rm ln}\left\{\frac{A}{T}\left[1-{\rm exp}\left(-\frac{2\alpha m_{\rm c}^\ast T}{H}\right)\right]\right\}
= -\frac{\alpha m_{\rm c}^\ast}{H}T+ const.,
\label{LK}
\end{equation}
where $A$ is the dHvA amplitude, $m_c^\ast$ is the cyclotron effective mass, and $\alpha = 2\pi^2 c k_{\rm B}/e\hbar$. The mass plot, namely, a plot of ${\rm ln}\{A[1-{\rm exp}(-2\alpha m_{\rm c}^\ast T/H)]/T\}$ against $T$, is shown in Fig.\ref{fig7}(b). The dHvA amplitudes of the $\alpha$, $\alpha'$, $\gamma_1$, and $\gamma_2$ branches are found to decay according to the LK formula, and the cyclotron effective masses were obtained from the slope of the fitting curves. The dHvA frequencies and the ratios of cyclotron effective mass to band mass are listed in Table \ref{t3}.
\begin{table}[t]
\caption{\label{t3}Experimental and calculated Fermi surface parameters of NbGe$_2$. $F$ and $F_{\rm b}$ are experimental and calculated dHvA frequencies, respectively. $m^\ast_{\rm c}$, $m_0$, and $m_{\rm b}$ are the cyclotron effective mass, rest mass, and band mass, respectively.}
\begin{ruledtabular}
\renewcommand{\arraystretch}{1.2}
\begin{tabular}{cccccc}
 & \multicolumn{2}{c}{Experiment} & \multicolumn{2}{c}{Calculation} & \\
\cline{2-5}
Branch & $F$ & $m^\ast_{\rm c}$ & $F_{\rm b}$ & $m_{\rm b}$ & $m^\ast_{\rm c}/m_{\rm b}$\\
& (kT) & ($m_0$) & (kT) & ($m_0$) & (-) \\
\hline
\multicolumn{6}{c}{$H$ $||$ $[0001]$}\\
 $\alpha$ & 11.7 & 1.89 & 11.0 & 1.23 & 1.54\\
 $\alpha'$ & 11.4 & 2.18 & 10.7 & 1.19 & 1.83\\
 $\gamma_1$ & 6.89 & 4.48 & 6.12 & 3.00 & 1.49\\
 $\gamma_2$ & 6.21 & 3.63 & 5.74 & 2.10 & 1.73\\
 $\zeta$ & 0.34 & 0.96 & 0.49 & 0.57 & 1.68 \\
 $\zeta'$ & 0.25 & 0.82  & 0.27 & 0.62 & 1.32\\
 $\beta^\ast$ & 4.07 & 3.07 & - & - & - \\
 $\kappa^\ast$ & 3.45 & 2.67 & - & - & - \\
\cline{1-6}
\multicolumn{6}{c}{$H$ $||$ $[10\bar{1}0]$}\\
 $\beta$ & 4.74 & 2.31 & 5.10 & 1.68 & 1.38\\
 $\beta'$ & 4.16 & 2.48 & 4.53 & 1.50 & 1.65\\
 $\gamma$ & 2.79 & 1.96 & 2.24 & 0.63 & 3.11\\
 $\gamma'$ & 3.00 & 2.25 & 2.36 & 1.19 & 1.89\\
 $\lambda_1$ & 0.14 & 0.35 & 0.21 & 0.27 & 1.30\\
 $\lambda_1'$ & 0.12 & 0.36 & 0.17 & 0.25 & 1.44\\
 $\eta^\ast$ & 2.25 & 1.40 & - & - & - \\
\cline{1-6}
\multicolumn{6}{c}{$H$ $||$ $[11\bar{2}0]$}\\
 $\beta^\ast$ & 4.26 & 2.21 & - & - & - \\
 $\gamma$ & 3.33 & 2.26 & 2.86 & 1.29 & 1.75\\
\end{tabular}
\end{ruledtabular}
\renewcommand{\arraystretch}{1}
\end{table}
The cyclotron effective masses of NbGe$_2$ were approximately 1.5 times larger than the band masses obtained by the LDA calculations, namely $m_{\rm c}^\ast$/$m_{\rm b}$ $\sim$ 1.5. Note that the calculated electronic specific heat coefficient $\gamma_{\rm calc}$ is obtained to be 4.3 mJ/(K$^2$ mole) from the LDA calculation. From the reported experimental value of $\gamma_{\rm exp} \sim 6.3$ mJ/(K$^2$ mole) \cite{BL20,EE20}, we estimated the ratio of $\gamma_{\rm exp}$/$\gamma_{\rm calc}$ to be approximately 1.5. The $m_{\rm c}^\ast$/$m_{\rm b}$ ratio is consistent with the $\gamma_{\rm exp}$/$\gamma_{\rm calc}$ ratio. 

Such mass enhancement is generally caused by the electronic correlations \cite{GRS84,LT88}, magnetic fluctuations/frustrations \cite{HW87,SK97,CU00}, and electron-phonon interactions \cite{IRG74,KM89}. NbGe$_2$ is a non-magnetic compound without $f$-electrons. Thus, the possibility that strong electronic correlations or magnetic contributions cause mass enhancement is ruled out. The results of dHvA measurements and first-principles calculations support that the strong electron-phonon interaction is responsible for the enhancement of the effective mass. Although the anisotropic electron-phonon scattering time has been demonstrated in the previous theoretical work \cite{CACG21}, rather isotropic effective mass enhancement was observed in NbGe$_2$ in this study.

As we have discussed, the Fermi surface of NbGe$_2$ is comprised of spin-split electron and spin-split hole Fermi surfaces. Such spin-split Fermi surfaces have been widely observed in non-centrosymmetric metals. We can estimate the energy scale of spin-splitting $\Delta \varepsilon$, namely the magnitude of anti-symmetric spin-orbit interaction, from the splitting of dHvA frequencies $\Delta F$ and the cyclotron effective mass $m_{\rm c}^\ast$. Using the Onsager relation $\Delta F = (c\hbar/2\pi e)\Delta S$ and the definition of the cyclotron mass $m_{\rm c}^\ast = (\hbar^2/2\pi)(\Delta S/\Delta \varepsilon)$, we obtain a simple expression as follows: 
\begin{equation}
\Delta \varepsilon = \frac{\hbar e}{m_{\rm c}^\ast c}\Delta F.
\label{ASOI}
\end{equation}
Note that the energy scale of the anti-symmetric spin-orbit interaction should be estimated using experimentally obtained effective masses in strongly correlated electron systems with enhanced effective masses \cite{VPM05,TT08,AM18,YJS20}. $\Delta \varepsilon$ of NbGe$_2$ is estimated to be 182 K for $\alpha$-$\alpha'$ branches (electron Fermi surface, $H$ $||$ $[0001]$) and 360 K for $\beta$-$\beta '$ branches (hole Fermi surface, $H$ $||$ $[10\bar{1}0]$). The estimated energy scale of spin-splitting is comparable to an isostructural compound NbSi$_2$ \cite{YO14} and a non-centrosymmetric metal LaRhGe$_3$ with Rh-4$d$ electrons \cite{TK08}.

Finally, let us compare the Fermi surface properties of NbGe$_2$ and NbSi$_2$. The Fermi surface properties of both compounds are summarized in Table$~$\ref{t4}.
\begin{table}[t]
\caption{\label{t4} Comparison of Fermi surface properties of NbGe$_2$ and NbSi$_2$. $\Delta F_{\rm exp}$ and $\Delta \varepsilon_{\rm exp}$ are the splitting of experimental dHvA frequencies and the corresponding energy scale of spin-splitting, respectively. The reported values of $\Delta F_{\rm exp}$, $\Delta \varepsilon_{\rm exp}$, $m_{\rm c}^\ast$, and $m_{\rm c}^\ast$/$m_{\rm b}$ of NbSi$_2$ are from ref.\cite{YO14}.}
\begin{ruledtabular}
\begin{tabular}{c c c c c }
\multicolumn{5}{c}{NbGe$_2$} \\ \hline
Branch & $\Delta F_{\rm exp}$ (T) & $\Delta \varepsilon_{\rm exp}$ (K) & $m_{\rm c}^\ast$ ($m_0$) & $m_{\rm c}^\ast$/$m_{\rm b}$ (-) \\ \hline
$\alpha$ \& $\alpha'$ & 270 & 182 & 1.89 \& 2.18 & 1.54 \& 1.83  \\
(electron) & & & &  \\
$\beta$ \& $\beta'$ & 586 & 360 & 2.31 \& 2.48 & 1.38 \& 1.65 \\ 
(hole) & & & &  \\ \hline
\multicolumn{5}{c}{NbSi$_2$ \cite{YO14}} \\ \hline
Branch & $\Delta F_{\rm exp}$ (T) & $\Delta \varepsilon_{\rm exp}$ (K) & $m_{\rm c}^\ast$ ($m_0$) & $m_{\rm c}^\ast$/$m_{\rm b}$ (-) \\ \hline
$\alpha$ \& $\alpha'$ & 120 & 209 & 0.73 \& 0.70 & 0.78 \& 0.78 \\
(electron) & & & &  \\
$\gamma$ \& $\gamma'$ & 240 & 334 & 0.98 \& 0.91 & 1.12 \& 1.10 \\ 
(hole) & & & & \\
\end{tabular}
\end{ruledtabular}
\end{table}
The Fermi surface of NbSi$_2$ has been investigated by dHvA experiments, and the hole Fermi surfaces consist of pocket Fermi surfaces around the K point in the Brillouin zone \cite{YO14}. In contrast, the (121st and 122nd) hole Fermi surfaces of NbGe$_2$ have a connected shape along the boundary of the Brillouin zone, as shown in Figs.$~$\ref{fig5}(c) and \ref{fig5}(d). The difference in the topology of hole Fermi surfaces as well as the density of states between NbGe$_2$ and NbSi$_2$ is arising from the different energy dispersion near the M point. As pointed out by the previous study \cite{EE20}, the energy band dispersion of NbGe$_2$ exhibits saddle-point-like dispersions near $E_{\rm F}$. Pairs of saddle-point-like dispersions exist at around the H and M points near $E_{\rm F}$ [see Fig.$~$\ref{fig6}], resulting in a distinct peak structure in the density of states. The van Hove-type singularity and the significant contribution of Nb-4$d$ electrons in the DOS near the Fermi level affect to the electronic transport and thermoelectric properties of NbGe$_2$. Note that the $\lambda_1$ and $\lambda_1'$ branches correspond to orbits circulating near the saddle-point at the M point. The  $m_{\rm c}^\ast$/$m_{\rm b}$ ratios for $\lambda_1$ and $\lambda_1'$ branches are comparable in magnitude to the mass enhancement observed in the other dHvA branches despite the presence of saddle-point-like band dispersion.
 
 The energy scales of spin-splitting $\Delta \varepsilon_{\rm exp}$ are comparable between NbGe$_2$ ($\Delta \varepsilon_{\rm exp} =$ 182 K for electron Fermi surfaces and $\Delta \varepsilon_{\rm exp} =$ 360 K for hole Fermi surfaces) and NbSi$_2$ ($\Delta \varepsilon_{\rm exp} =$ 209 K for electron Fermi surfaces and $\Delta \varepsilon_{\rm exp} =$ 334 K for hole Fermi surfaces \cite{YO14}). In NbGe$_2$, the effective mass is enhanced by the strong electron-phonon interaction compared to NbSi2. As a result, while the spin-splitting of dHvA frequencies $\Delta F$ differs significantly between NbGe$_2$ and NbSi$_2$, the energy scale of spin-splitting remains comparable.

\section{Conclusions}
We investigated the detailed Fermi surface topology and electronic transport properties of NbGe$_2$. NbGe$_2$ and NbSi$_2$ exhibit a notable difference in the temperature dependence of electrical resistivity and Seebeck coefficient due to the strong interband electron-phonon scattering. The characteristic behavior in $\rho(T)$ and $S(T)$ of NbGe$_2$ are attributed to the van Hove-type peak structure and the significant contribution of Nb-4$d$ electrons in the DOS near the $E_{\rm F}$. The detailed topology of the Fermi surfaces in NbGe$_2$ is revealed by combining the magnetic field modulation and magnetic torque methods. The angular dependence of the experimental dHvA frequencies can be interpreted based on the results of LDA calculations. The dHvA measurements also revealed the isotropic enhancement of the cyclotron effective mass ($m_{\rm c}^\ast/m_{\rm b} \sim 1.5$), indicating the existence of strong electron-phonon interaction in NbGe$_2$. Our results provide valuable insights into the intriguing electronic structure of NbGe$_2$, which is a candidate for a coupled electron-phonon liquid.

\begin{acknowledgments}
We would like to thank Y. Haga, Y \={O}nuki, N. Kimura, F. Honda, Y. Shimizu, Y. Homma, and D. X. Li for discussion. We acknowledge all the support from the International Research Center for Nuclear Materials Science at Oarai (Institute for Materials Research, Tohoku University). This work was supported by KAKENHI (JP22K20360, JP22H04933, JP21K03448), Grant-in-Aid for JSPS Research Fellow (JP19J20539), and DIARE research grant.
\end{acknowledgments}

\bibliography{NbGe2}

\providecommand{\noopsort}[1]{}\providecommand{\singleletter}[1]{#1}%
\begin{thebibliography}{47}%
\makeatletter
\providecommand \@ifxundefined [1]{%
 \@ifx{#1\undefined}
}%
\providecommand \@ifnum [1]{%
 \ifnum #1\expandafter \@firstoftwo
 \else \expandafter \@secondoftwo
 \fi
}%
\providecommand \@ifx [1]{%
 \ifx #1\expandafter \@firstoftwo
 \else \expandafter \@secondoftwo
 \fi
}%
\providecommand \natexlab [1]{#1}%
\providecommand \enquote  [1]{``#1''}%
\providecommand \bibnamefont  [1]{#1}%
\providecommand \bibfnamefont [1]{#1}%
\providecommand \citenamefont [1]{#1}%
\providecommand \href@noop [0]{\@secondoftwo}%
\providecommand \href [0]{\begingroup \@sanitize@url \@href}%
\providecommand \@href[1]{\@@startlink{#1}\@@href}%
\providecommand \@@href[1]{\endgroup#1\@@endlink}%
\providecommand \@sanitize@url [0]{\catcode `\\12\catcode `\$12\catcode
  `\&12\catcode `\#12\catcode `\^12\catcode `\_12\catcode `\%12\relax}%
\providecommand \@@startlink[1]{}%
\providecommand \@@endlink[0]{}%
\providecommand \url  [0]{\begingroup\@sanitize@url \@url }%
\providecommand \@url [1]{\endgroup\@href {#1}{\urlprefix }}%
\providecommand \urlprefix  [0]{URL }%
\providecommand \Eprint [0]{\href }%
\providecommand \doibase [0]{http://dx.doi.org/}%
\providecommand \selectlanguage [0]{\@gobble}%
\providecommand \bibinfo  [0]{\@secondoftwo}%
\providecommand \bibfield  [0]{\@secondoftwo}%
\providecommand \translation [1]{[#1]}%
\providecommand \BibitemOpen [0]{}%
\providecommand \bibitemStop [0]{}%
\providecommand \bibitemNoStop [0]{.\EOS\space}%
\providecommand \EOS [0]{\spacefactor3000\relax}%
\providecommand \BibitemShut  [1]{\csname bibitem#1\endcsname}%
\let\auto@bib@innerbib\@empty
\bibitem [{\citenamefont {Rikken}\ and\ \citenamefont
  {Raupach}(1997)}]{GLJAR97}%
  \BibitemOpen
  \bibfield  {author} {\bibinfo {author} {\bibfnamefont {G.~L. J.~A.}\
  \bibnamefont {Rikken}}\ and\ \bibinfo {author} {\bibfnamefont
  {E.}~\bibnamefont {Raupach}},\ }\href@noop {} {\bibfield  {journal} {\bibinfo
   {journal} {Nature}\ }\textbf {\bibinfo {volume} {390}},\ \bibinfo {pages}
  {493} (\bibinfo {year} {1997})}\BibitemShut {NoStop}%
\bibitem [{\citenamefont {M{\"{u}}hlbauer}\ \emph {et~al.}(2009)\citenamefont
  {M{\"{u}}hlbauer}, \citenamefont {Binz}, \citenamefont {Jonietz},
  \citenamefont {Pfleiderer}, \citenamefont {Rosch}, \citenamefont {Neubauer},
  \citenamefont {Georgii},\ and\ \citenamefont {B{\"{o}}ni}}]{SM09}%
  \BibitemOpen
  \bibfield  {author} {\bibinfo {author} {\bibfnamefont {S.}~\bibnamefont
  {M{\"{u}}hlbauer}}, \bibinfo {author} {\bibfnamefont {B.}~\bibnamefont
  {Binz}}, \bibinfo {author} {\bibfnamefont {F.}~\bibnamefont {Jonietz}},
  \bibinfo {author} {\bibfnamefont {C.}~\bibnamefont {Pfleiderer}}, \bibinfo
  {author} {\bibfnamefont {A.}~\bibnamefont {Rosch}}, \bibinfo {author}
  {\bibfnamefont {A.}~\bibnamefont {Neubauer}}, \bibinfo {author}
  {\bibfnamefont {R.}~\bibnamefont {Georgii}}, \ and\ \bibinfo {author}
  {\bibfnamefont {P.}~\bibnamefont {B{\"{o}}ni}},\ }\href@noop {} {\bibfield
  {journal} {\bibinfo  {journal} {Science}\ }\textbf {\bibinfo {volume}
  {\textbf{323}}},\ \bibinfo {pages} {915} (\bibinfo {year}
  {2009})}\BibitemShut {NoStop}%
\bibitem [{\citenamefont {Togawa}\ \emph {et~al.}(2012)\citenamefont {Togawa},
  \citenamefont {Koyama}, \citenamefont {Takayanagi}, \citenamefont {Mori},
  \citenamefont {Kousaka}, \citenamefont {Akimitsu}, \citenamefont {Nishihara},
  \citenamefont {Inoue}, \citenamefont {Ovchinnikov},\ and\ \citenamefont
  {Kishine}}]{YT12}%
  \BibitemOpen
  \bibfield  {author} {\bibinfo {author} {\bibfnamefont {Y.}~\bibnamefont
  {Togawa}}, \bibinfo {author} {\bibfnamefont {T.}~\bibnamefont {Koyama}},
  \bibinfo {author} {\bibfnamefont {K.}~\bibnamefont {Takayanagi}}, \bibinfo
  {author} {\bibfnamefont {S.}~\bibnamefont {Mori}}, \bibinfo {author}
  {\bibfnamefont {Y.}~\bibnamefont {Kousaka}}, \bibinfo {author} {\bibfnamefont
  {J.}~\bibnamefont {Akimitsu}}, \bibinfo {author} {\bibfnamefont
  {S.}~\bibnamefont {Nishihara}}, \bibinfo {author} {\bibfnamefont
  {K.}~\bibnamefont {Inoue}}, \bibinfo {author} {\bibfnamefont {A.~S.}\
  \bibnamefont {Ovchinnikov}}, \ and\ \bibinfo {author} {\bibfnamefont
  {J.}~\bibnamefont {Kishine}},\ }\href@noop {} {\bibfield  {journal} {\bibinfo
   {journal} {Phys.\ Rev.\ Lett.}\ }\textbf {\bibinfo {volume}
  {\textbf{108}}},\ \bibinfo {pages} {107202} (\bibinfo {year}
  {2012})}\BibitemShut {NoStop}%
\bibitem [{\citenamefont {Sato}\ \emph {et~al.}(2022)\citenamefont {Sato},
  \citenamefont {Manako}, \citenamefont {Homma}, \citenamefont {Li},
  \citenamefont {Okazaki},\ and\ \citenamefont {Aoki}}]{YJS22}%
  \BibitemOpen
  \bibfield  {author} {\bibinfo {author} {\bibfnamefont {Y.~J.}\ \bibnamefont
  {Sato}}, \bibinfo {author} {\bibfnamefont {H.}~\bibnamefont {Manako}},
  \bibinfo {author} {\bibfnamefont {Y.}~\bibnamefont {Homma}}, \bibinfo
  {author} {\bibfnamefont {D.~X.}\ \bibnamefont {Li}}, \bibinfo {author}
  {\bibfnamefont {R.}~\bibnamefont {Okazaki}}, \ and\ \bibinfo {author}
  {\bibfnamefont {D.}~\bibnamefont {Aoki}},\ }\href@noop {} {\bibfield
  {journal} {\bibinfo  {journal} {Phys.\ Rev.\ Mater.}\ }\textbf {\bibinfo
  {volume} {\textbf{6}}},\ \bibinfo {pages} {104412} (\bibinfo {year}
  {2022})}\BibitemShut {NoStop}%
\bibitem [{\citenamefont {Sato}\ \emph {et~al.}(2023)\citenamefont {Sato},
  \citenamefont {Manako}, \citenamefont {Okazaki}, \citenamefont {Yasui},
  \citenamefont {Nakamura},\ and\ \citenamefont {Aoki}}]{YJS23}%
  \BibitemOpen
  \bibfield  {author} {\bibinfo {author} {\bibfnamefont {Y.~J.}\ \bibnamefont
  {Sato}}, \bibinfo {author} {\bibfnamefont {H.}~\bibnamefont {Manako}},
  \bibinfo {author} {\bibfnamefont {R.}~\bibnamefont {Okazaki}}, \bibinfo
  {author} {\bibfnamefont {Y.}~\bibnamefont {Yasui}}, \bibinfo {author}
  {\bibfnamefont {A.}~\bibnamefont {Nakamura}}, \ and\ \bibinfo {author}
  {\bibfnamefont {D.}~\bibnamefont {Aoki}},\ }\href@noop {} {\bibfield
  {journal} {\bibinfo  {journal} {Phys.\ Rev.\ B}\ }\textbf {\bibinfo {volume}
  {\textbf{107}}},\ \bibinfo {pages} {214420} (\bibinfo {year}
  {2023})}\BibitemShut {NoStop}%
\bibitem [{\citenamefont {Rikken}\ \emph {et~al.}(2001)\citenamefont {Rikken},
  \citenamefont {F{\"{o}}lling},\ and\ \citenamefont {Wyder}}]{GLJAR01}%
  \BibitemOpen
  \bibfield  {author} {\bibinfo {author} {\bibfnamefont {G.~L. J.~A.}\
  \bibnamefont {Rikken}}, \bibinfo {author} {\bibfnamefont {J.}~\bibnamefont
  {F{\"{o}}lling}}, \ and\ \bibinfo {author} {\bibfnamefont {P.}~\bibnamefont
  {Wyder}},\ }\href@noop {} {\bibfield  {journal} {\bibinfo  {journal} {Phys.\
  Rev.\ Lett.}\ }\textbf {\bibinfo {volume} {87}},\ \bibinfo {pages} {236602}
  (\bibinfo {year} {2001})}\BibitemShut {NoStop}%
\bibitem [{\citenamefont {Tsirkin}\ \emph {et~al.}(2017)\citenamefont
  {Tsirkin}, \citenamefont {Souza},\ and\ \citenamefont {Vanderbilt}}]{SST17}%
  \BibitemOpen
  \bibfield  {author} {\bibinfo {author} {\bibfnamefont {S.~S.}\ \bibnamefont
  {Tsirkin}}, \bibinfo {author} {\bibfnamefont {I.}~\bibnamefont {Souza}}, \
  and\ \bibinfo {author} {\bibfnamefont {D.}~\bibnamefont {Vanderbilt}},\
  }\href@noop {} {\bibfield  {journal} {\bibinfo  {journal} {Phys.\ Rev.\ B}\
  }\textbf {\bibinfo {volume} {96}},\ \bibinfo {pages} {045102} (\bibinfo
  {year} {2017})}\BibitemShut {NoStop}%
\bibitem [{\citenamefont {Chang}\ \emph {et~al.}(2018)\citenamefont {Chang},
  \citenamefont {Wieder}, \citenamefont {Schindler}, \citenamefont {Sanchez},
  \citenamefont {Belopolski}, \citenamefont {Huang}, \citenamefont {Singh},
  \citenamefont {Wu}, \citenamefont {Chang}, \citenamefont {Neupert},
  \citenamefont {Xu}, \citenamefont {Lin},\ and\ \citenamefont {Hasan}}]{GC18}%
  \BibitemOpen
  \bibfield  {author} {\bibinfo {author} {\bibfnamefont {G.}~\bibnamefont
  {Chang}}, \bibinfo {author} {\bibfnamefont {B.~J.}\ \bibnamefont {Wieder}},
  \bibinfo {author} {\bibfnamefont {F.}~\bibnamefont {Schindler}}, \bibinfo
  {author} {\bibfnamefont {D.~S.}\ \bibnamefont {Sanchez}}, \bibinfo {author}
  {\bibfnamefont {I.}~\bibnamefont {Belopolski}}, \bibinfo {author}
  {\bibfnamefont {S.-M.}\ \bibnamefont {Huang}}, \bibinfo {author}
  {\bibfnamefont {B.}~\bibnamefont {Singh}}, \bibinfo {author} {\bibfnamefont
  {D.}~\bibnamefont {Wu}}, \bibinfo {author} {\bibfnamefont {T.-R.}\
  \bibnamefont {Chang}}, \bibinfo {author} {\bibfnamefont {T.}~\bibnamefont
  {Neupert}}, \bibinfo {author} {\bibfnamefont {S.-Y.}\ \bibnamefont {Xu}},
  \bibinfo {author} {\bibfnamefont {H.}~\bibnamefont {Lin}}, \ and\ \bibinfo
  {author} {\bibfnamefont {M.~Z.}\ \bibnamefont {Hasan}},\ }\href@noop {}
  {\bibfield  {journal} {\bibinfo  {journal} {Nat.\ Mater.}\ }\textbf {\bibinfo
  {volume} {17}},\ \bibinfo {pages} {978} (\bibinfo {year} {2018})}\BibitemShut
  {NoStop}%
\bibitem [{\citenamefont {Chang}\ \emph {et~al.}(2017)\citenamefont {Chang},
  \citenamefont {Xu}, \citenamefont {Wieder}, \citenamefont {Sanchez},
  \citenamefont {Huang}, \citenamefont {Belopolski}, \citenamefont {Chang},
  \citenamefont {Zhang}, \citenamefont {Bansil}, \citenamefont {Lin},\ and\
  \citenamefont {Hasan}}]{GC17}%
  \BibitemOpen
  \bibfield  {author} {\bibinfo {author} {\bibfnamefont {G.}~\bibnamefont
  {Chang}}, \bibinfo {author} {\bibfnamefont {S.-Y.}\ \bibnamefont {Xu}},
  \bibinfo {author} {\bibfnamefont {B.~J.}\ \bibnamefont {Wieder}}, \bibinfo
  {author} {\bibfnamefont {D.~S.}\ \bibnamefont {Sanchez}}, \bibinfo {author}
  {\bibfnamefont {S.-M.}\ \bibnamefont {Huang}}, \bibinfo {author}
  {\bibfnamefont {I.}~\bibnamefont {Belopolski}}, \bibinfo {author}
  {\bibfnamefont {T.-R.}\ \bibnamefont {Chang}}, \bibinfo {author}
  {\bibfnamefont {S.}~\bibnamefont {Zhang}}, \bibinfo {author} {\bibfnamefont
  {A.}~\bibnamefont {Bansil}}, \bibinfo {author} {\bibfnamefont
  {H.}~\bibnamefont {Lin}}, \ and\ \bibinfo {author} {\bibfnamefont {M.~Z.}\
  \bibnamefont {Hasan}},\ }\href@noop {} {\bibfield  {journal} {\bibinfo
  {journal} {Phys.\ Rev.\ Lett.}\ }\textbf {\bibinfo {volume} {119}},\ \bibinfo
  {pages} {206401} (\bibinfo {year} {2017})}\BibitemShut {NoStop}%
\bibitem [{\citenamefont {Tang}\ \emph {et~al.}(2017)\citenamefont {Tang},
  \citenamefont {Zhou},\ and\ \citenamefont {Zhang}}]{PT17}%
  \BibitemOpen
  \bibfield  {author} {\bibinfo {author} {\bibfnamefont {P.}~\bibnamefont
  {Tang}}, \bibinfo {author} {\bibfnamefont {Q.}~\bibnamefont {Zhou}}, \ and\
  \bibinfo {author} {\bibfnamefont {S.-C.}\ \bibnamefont {Zhang}},\ }\href@noop
  {} {\bibfield  {journal} {\bibinfo  {journal} {Phys.\ Rev.\ Lett.}\ }\textbf
  {\bibinfo {volume} {119}},\ \bibinfo {pages} {206402} (\bibinfo {year}
  {2017})}\BibitemShut {NoStop}%
\bibitem [{\citenamefont {Zhang}\ \emph {et~al.}(2018)\citenamefont {Zhang},
  \citenamefont {Song}, \citenamefont {Alexandradinata}, \citenamefont {Weng},
  \citenamefont {Fang}, \citenamefont {Lu},\ and\ \citenamefont {Fang}}]{TZ18}%
  \BibitemOpen
  \bibfield  {author} {\bibinfo {author} {\bibfnamefont {T.}~\bibnamefont
  {Zhang}}, \bibinfo {author} {\bibfnamefont {Z.}~\bibnamefont {Song}},
  \bibinfo {author} {\bibfnamefont {A.}~\bibnamefont {Alexandradinata}},
  \bibinfo {author} {\bibfnamefont {H.}~\bibnamefont {Weng}}, \bibinfo {author}
  {\bibfnamefont {C.}~\bibnamefont {Fang}}, \bibinfo {author} {\bibfnamefont
  {L.}~\bibnamefont {Lu}}, \ and\ \bibinfo {author} {\bibfnamefont
  {Z.}~\bibnamefont {Fang}},\ }\href@noop {} {\bibfield  {journal} {\bibinfo
  {journal} {Phys.\ Rev.\ Lett.}\ }\textbf {\bibinfo {volume} {120}},\ \bibinfo
  {pages} {016401} (\bibinfo {year} {2018})}\BibitemShut {NoStop}%
\bibitem [{\citenamefont {Furukawa}\ \emph {et~al.}(2017)\citenamefont
  {Furukawa}, \citenamefont {Shimokawa}, \citenamefont {Kobayashi},\ and\
  \citenamefont {Itou}}]{TF17}%
  \BibitemOpen
  \bibfield  {author} {\bibinfo {author} {\bibfnamefont {T.}~\bibnamefont
  {Furukawa}}, \bibinfo {author} {\bibfnamefont {Y.}~\bibnamefont {Shimokawa}},
  \bibinfo {author} {\bibfnamefont {K.}~\bibnamefont {Kobayashi}}, \ and\
  \bibinfo {author} {\bibfnamefont {T.}~\bibnamefont {Itou}},\ }\href@noop {}
  {\bibfield  {journal} {\bibinfo  {journal} {Nat.\ Commun.}\ }\textbf
  {\bibinfo {volume} {8}},\ \bibinfo {pages} {954} (\bibinfo {year}
  {2017})}\BibitemShut {NoStop}%
\bibitem [{\citenamefont {Sakano}\ \emph {et~al.}(2020)\citenamefont {Sakano},
  \citenamefont {Hirayama}, \citenamefont {Takahashi}, \citenamefont {Akebi},
  \citenamefont {Nakayama}, \citenamefont {Kuroda}, \citenamefont {Taguchi},
  \citenamefont {Yoshikawa}, \citenamefont {Miyamoto}, \citenamefont {Okuda},
  \citenamefont {Ono}, \citenamefont {Kumigashira}, \citenamefont {Ideue},
  \citenamefont {Iwasa}, \citenamefont {Mitsuishi}, \citenamefont {Ishizaka},
  \citenamefont {Shin}, \citenamefont {Miyake}, \citenamefont {Murakami},
  \citenamefont {Sasagawa},\ and\ \citenamefont {Kondo}}]{MS20}%
  \BibitemOpen
  \bibfield  {author} {\bibinfo {author} {\bibfnamefont {M.}~\bibnamefont
  {Sakano}}, \bibinfo {author} {\bibfnamefont {M.}~\bibnamefont {Hirayama}},
  \bibinfo {author} {\bibfnamefont {T.}~\bibnamefont {Takahashi}}, \bibinfo
  {author} {\bibfnamefont {S.}~\bibnamefont {Akebi}}, \bibinfo {author}
  {\bibfnamefont {M.}~\bibnamefont {Nakayama}}, \bibinfo {author}
  {\bibfnamefont {K.}~\bibnamefont {Kuroda}}, \bibinfo {author} {\bibfnamefont
  {K.}~\bibnamefont {Taguchi}}, \bibinfo {author} {\bibfnamefont
  {T.}~\bibnamefont {Yoshikawa}}, \bibinfo {author} {\bibfnamefont
  {K.}~\bibnamefont {Miyamoto}}, \bibinfo {author} {\bibfnamefont
  {T.}~\bibnamefont {Okuda}}, \bibinfo {author} {\bibfnamefont
  {K.}~\bibnamefont {Ono}}, \bibinfo {author} {\bibfnamefont {H.}~\bibnamefont
  {Kumigashira}}, \bibinfo {author} {\bibfnamefont {T.}~\bibnamefont {Ideue}},
  \bibinfo {author} {\bibfnamefont {Y.}~\bibnamefont {Iwasa}}, \bibinfo
  {author} {\bibfnamefont {N.}~\bibnamefont {Mitsuishi}}, \bibinfo {author}
  {\bibfnamefont {K.}~\bibnamefont {Ishizaka}}, \bibinfo {author}
  {\bibfnamefont {S.}~\bibnamefont {Shin}}, \bibinfo {author} {\bibfnamefont
  {T.}~\bibnamefont {Miyake}}, \bibinfo {author} {\bibfnamefont
  {S.}~\bibnamefont {Murakami}}, \bibinfo {author} {\bibfnamefont
  {T.}~\bibnamefont {Sasagawa}}, \ and\ \bibinfo {author} {\bibfnamefont
  {T.}~\bibnamefont {Kondo}},\ }\href@noop {} {\bibfield  {journal} {\bibinfo
  {journal} {Phys.\ Rev.\ Lett.}\ }\textbf {\bibinfo {volume} {124}},\ \bibinfo
  {pages} {136404} (\bibinfo {year} {2020})}\BibitemShut {NoStop}%
\bibitem [{\citenamefont {Gatti}\ \emph {et~al.}(2020)\citenamefont {Gatti},
  \citenamefont {Gos\'{a}lbez-Mart\'{i}nez}, \citenamefont {Tsirkin},
  \citenamefont {Fanciulli}, \citenamefont {Puppin}, \citenamefont
  {Polishchuk}, \citenamefont {Moser}, \citenamefont {Testa}, \citenamefont
  {Martino}, \citenamefont {Roth}, \citenamefont {Bugnon}, \citenamefont
  {Moreschini}, \citenamefont {Bostwick}, \citenamefont {Jozwiak},
  \citenamefont {Rotenberg}, \citenamefont {DiSanto}, \citenamefont {Petaccia},
  \citenamefont {Vobornik}, \citenamefont {Fujii}, \citenamefont {Wong},
  \citenamefont {Jariwala}, \citenamefont {Atwater}, \citenamefont {R{\o}nnow},
  \citenamefont {Chergui}, \citenamefont {Yazyev}, \citenamefont {Grioni},\
  and\ \citenamefont {Crepaldi}}]{GG20}%
  \BibitemOpen
  \bibfield  {author} {\bibinfo {author} {\bibfnamefont {G.}~\bibnamefont
  {Gatti}}, \bibinfo {author} {\bibfnamefont {D.}~\bibnamefont
  {Gos\'{a}lbez-Mart\'{i}nez}}, \bibinfo {author} {\bibfnamefont {S.~S.}\
  \bibnamefont {Tsirkin}}, \bibinfo {author} {\bibfnamefont {M.}~\bibnamefont
  {Fanciulli}}, \bibinfo {author} {\bibfnamefont {M.}~\bibnamefont {Puppin}},
  \bibinfo {author} {\bibfnamefont {S.}~\bibnamefont {Polishchuk}}, \bibinfo
  {author} {\bibfnamefont {S.}~\bibnamefont {Moser}}, \bibinfo {author}
  {\bibfnamefont {L.}~\bibnamefont {Testa}}, \bibinfo {author} {\bibfnamefont
  {E.}~\bibnamefont {Martino}}, \bibinfo {author} {\bibfnamefont
  {S.}~\bibnamefont {Roth}}, \bibinfo {author} {\bibfnamefont {P.}~\bibnamefont
  {Bugnon}}, \bibinfo {author} {\bibfnamefont {L.}~\bibnamefont {Moreschini}},
  \bibinfo {author} {\bibfnamefont {A.}~\bibnamefont {Bostwick}}, \bibinfo
  {author} {\bibfnamefont {C.}~\bibnamefont {Jozwiak}}, \bibinfo {author}
  {\bibfnamefont {E.}~\bibnamefont {Rotenberg}}, \bibinfo {author}
  {\bibfnamefont {G.}~\bibnamefont {DiSanto}}, \bibinfo {author} {\bibfnamefont
  {L.}~\bibnamefont {Petaccia}}, \bibinfo {author} {\bibfnamefont
  {I.}~\bibnamefont {Vobornik}}, \bibinfo {author} {\bibfnamefont
  {J.}~\bibnamefont {Fujii}}, \bibinfo {author} {\bibfnamefont
  {J.}~\bibnamefont {Wong}}, \bibinfo {author} {\bibfnamefont {D.}~\bibnamefont
  {Jariwala}}, \bibinfo {author} {\bibfnamefont {H.~A.}\ \bibnamefont
  {Atwater}}, \bibinfo {author} {\bibfnamefont {H.~M.}\ \bibnamefont
  {R{\o}nnow}}, \bibinfo {author} {\bibfnamefont {M.}~\bibnamefont {Chergui}},
  \bibinfo {author} {\bibfnamefont {O.~V.}\ \bibnamefont {Yazyev}}, \bibinfo
  {author} {\bibfnamefont {M.}~\bibnamefont {Grioni}}, \ and\ \bibinfo {author}
  {\bibfnamefont {A.}~\bibnamefont {Crepaldi}},\ }\href@noop {} {\bibfield
  {journal} {\bibinfo  {journal} {Phys.\ Rev.\ Lett.}\ }\textbf {\bibinfo
  {volume} {125}},\ \bibinfo {pages} {216402} (\bibinfo {year}
  {2020})}\BibitemShut {NoStop}%
\bibitem [{\citenamefont {Remeika}\ \emph {et~al.}(1978)\citenamefont
  {Remeika}, \citenamefont {Cooper}, \citenamefont {Fisk},\ and\ \citenamefont
  {Johnston}}]{JPR78}%
  \BibitemOpen
  \bibfield  {author} {\bibinfo {author} {\bibfnamefont {J.~P.}\ \bibnamefont
  {Remeika}}, \bibinfo {author} {\bibfnamefont {A.~S.}\ \bibnamefont {Cooper}},
  \bibinfo {author} {\bibfnamefont {Z.}~\bibnamefont {Fisk}}, \ and\ \bibinfo
  {author} {\bibfnamefont {D.~C.}\ \bibnamefont {Johnston}},\ }\href@noop {}
  {\bibfield  {journal} {\bibinfo  {journal} {J.\ Less-Common\ Met.}\ }\textbf
  {\bibinfo {volume} {62}},\ \bibinfo {pages} {211} (\bibinfo {year}
  {1978})}\BibitemShut {NoStop}%
\bibitem [{\citenamefont {Lv}\ \emph {et~al.}(2020)\citenamefont {Lv},
  \citenamefont {Li}, \citenamefont {Chen}, \citenamefont {Yang}, \citenamefont
  {Wu}, \citenamefont {Qiao}, \citenamefont {Guan}, \citenamefont {Xing},
  \citenamefont {Tao}, \citenamefont {Cao},\ and\ \citenamefont {Xu}}]{BL20}%
  \BibitemOpen
  \bibfield  {author} {\bibinfo {author} {\bibfnamefont {B.}~\bibnamefont
  {Lv}}, \bibinfo {author} {\bibfnamefont {M.}~\bibnamefont {Li}}, \bibinfo
  {author} {\bibfnamefont {J.}~\bibnamefont {Chen}}, \bibinfo {author}
  {\bibfnamefont {Y.}~\bibnamefont {Yang}}, \bibinfo {author} {\bibfnamefont
  {S.}~\bibnamefont {Wu}}, \bibinfo {author} {\bibfnamefont {L.}~\bibnamefont
  {Qiao}}, \bibinfo {author} {\bibfnamefont {F.}~\bibnamefont {Guan}}, \bibinfo
  {author} {\bibfnamefont {H.}~\bibnamefont {Xing}}, \bibinfo {author}
  {\bibfnamefont {Q.}~\bibnamefont {Tao}}, \bibinfo {author} {\bibfnamefont
  {G.-H.}\ \bibnamefont {Cao}}, \ and\ \bibinfo {author} {\bibfnamefont
  {Z.-A.}\ \bibnamefont {Xu}},\ }\href@noop {} {\bibfield  {journal} {\bibinfo
  {journal} {Phys.\ Rev.\ B}\ }\textbf {\bibinfo {volume} {102}},\ \bibinfo
  {pages} {064507} (\bibinfo {year} {2020})}\BibitemShut {NoStop}%
\bibitem [{\citenamefont {Emmanouilidou}\ \emph {et~al.}(2020)\citenamefont
  {Emmanouilidou}, \citenamefont {Mardanya}, \citenamefont {Xing},
  \citenamefont {Reddy}, \citenamefont {Agarwal}, \citenamefont {Chang},\ and\
  \citenamefont {Ni}}]{EE20}%
  \BibitemOpen
  \bibfield  {author} {\bibinfo {author} {\bibfnamefont {E.}~\bibnamefont
  {Emmanouilidou}}, \bibinfo {author} {\bibfnamefont {S.}~\bibnamefont
  {Mardanya}}, \bibinfo {author} {\bibfnamefont {J.}~\bibnamefont {Xing}},
  \bibinfo {author} {\bibfnamefont {P.~V.~S.}\ \bibnamefont {Reddy}}, \bibinfo
  {author} {\bibfnamefont {A.}~\bibnamefont {Agarwal}}, \bibinfo {author}
  {\bibfnamefont {T.-R.}\ \bibnamefont {Chang}}, \ and\ \bibinfo {author}
  {\bibfnamefont {N.}~\bibnamefont {Ni}},\ }\href@noop {} {\bibfield  {journal}
  {\bibinfo  {journal} {Phys.\ Rev.\ B}\ }\textbf {\bibinfo {volume} {102}},\
  \bibinfo {pages} {235144} (\bibinfo {year} {2020})}\BibitemShut {NoStop}%
\bibitem [{\citenamefont {Zhang}\ \emph {et~al.}(2021)\citenamefont {Zhang},
  \citenamefont {Le}, \citenamefont {Lv.}, \citenamefont {Yin}, \citenamefont
  {Chen}, \citenamefont {Nie}, \citenamefont {Su}, \citenamefont {Yuan},
  \citenamefont {Xu},\ and\ \citenamefont {Lu}}]{DZ21}%
  \BibitemOpen
  \bibfield  {author} {\bibinfo {author} {\bibfnamefont {D.}~\bibnamefont
  {Zhang}}, \bibinfo {author} {\bibfnamefont {T.}~\bibnamefont {Le}}, \bibinfo
  {author} {\bibfnamefont {B.}~\bibnamefont {Lv.}}, \bibinfo {author}
  {\bibfnamefont {L.}~\bibnamefont {Yin}}, \bibinfo {author} {\bibfnamefont
  {C.}~\bibnamefont {Chen}}, \bibinfo {author} {\bibfnamefont {Z.}~\bibnamefont
  {Nie}}, \bibinfo {author} {\bibfnamefont {D.}~\bibnamefont {Su}}, \bibinfo
  {author} {\bibfnamefont {H.}~\bibnamefont {Yuan}}, \bibinfo {author}
  {\bibfnamefont {Z.-A.}\ \bibnamefont {Xu}}, \ and\ \bibinfo {author}
  {\bibfnamefont {X.}~\bibnamefont {Lu}},\ }\href@noop {} {\bibfield  {journal}
  {\bibinfo  {journal} {Phys.\ Rev.\ B}\ }\textbf {\bibinfo {volume} {103}},\
  \bibinfo {pages} {214508} (\bibinfo {year} {2021})}\BibitemShut {NoStop}%
\bibitem [{\citenamefont {Garcia}\ \emph {et~al.}(2021)\citenamefont {Garcia},
  \citenamefont {Nenno}, \citenamefont {Varnavides},\ and\ \citenamefont
  {Narang}}]{CACG21}%
  \BibitemOpen
  \bibfield  {author} {\bibinfo {author} {\bibfnamefont {C.~A.~C.}\
  \bibnamefont {Garcia}}, \bibinfo {author} {\bibfnamefont {D.~M.}\
  \bibnamefont {Nenno}}, \bibinfo {author} {\bibfnamefont {G.}~\bibnamefont
  {Varnavides}}, \ and\ \bibinfo {author} {\bibfnamefont {P.}~\bibnamefont
  {Narang}},\ }\href@noop {} {\bibfield  {journal} {\bibinfo  {journal} {Phys.\
  Rev.\ Mater.}\ }\textbf {\bibinfo {volume} {5}},\ \bibinfo {pages} {L091202}
  (\bibinfo {year} {2021})}\BibitemShut {NoStop}%
\bibitem [{\citenamefont {Yang}\ \emph {et~al.}(2021)\citenamefont {Yang},
  \citenamefont {Yao}, \citenamefont {Plisson}, \citenamefont {Mozaffari},
  \citenamefont {Scheifers}, \citenamefont {Savvidou}, \citenamefont {Choi},
  \citenamefont {McCandless}, \citenamefont {Padlewski}, \citenamefont
  {Putzke}, \citenamefont {Moll}, \citenamefont {Chan}, \citenamefont
  {Balicas}, \citenamefont {Burch},\ and\ \citenamefont {Tafti}}]{HYY21}%
  \BibitemOpen
  \bibfield  {author} {\bibinfo {author} {\bibfnamefont {H.-Y.}\ \bibnamefont
  {Yang}}, \bibinfo {author} {\bibfnamefont {X.}~\bibnamefont {Yao}}, \bibinfo
  {author} {\bibfnamefont {V.}~\bibnamefont {Plisson}}, \bibinfo {author}
  {\bibfnamefont {S.}~\bibnamefont {Mozaffari}}, \bibinfo {author}
  {\bibfnamefont {J.~P.}\ \bibnamefont {Scheifers}}, \bibinfo {author}
  {\bibfnamefont {A.~F.}\ \bibnamefont {Savvidou}}, \bibinfo {author}
  {\bibfnamefont {E.~S.}\ \bibnamefont {Choi}}, \bibinfo {author}
  {\bibfnamefont {G.~T.}\ \bibnamefont {McCandless}}, \bibinfo {author}
  {\bibfnamefont {M.~F.}\ \bibnamefont {Padlewski}}, \bibinfo {author}
  {\bibfnamefont {C.}~\bibnamefont {Putzke}}, \bibinfo {author} {\bibfnamefont
  {P.~J.~W.}\ \bibnamefont {Moll}}, \bibinfo {author} {\bibfnamefont {J.~Y.}\
  \bibnamefont {Chan}}, \bibinfo {author} {\bibfnamefont {L.}~\bibnamefont
  {Balicas}}, \bibinfo {author} {\bibfnamefont {K.~S.}\ \bibnamefont {Burch}},
  \ and\ \bibinfo {author} {\bibfnamefont {F.}~\bibnamefont {Tafti}},\
  }\href@noop {} {\bibfield  {journal} {\bibinfo  {journal} {Nat.\ Commun.}\
  }\textbf {\bibinfo {volume} {12}},\ \bibinfo {pages} {5292} (\bibinfo {year}
  {2021})}\BibitemShut {NoStop}%
\bibitem [{\citenamefont {Levchenko}\ and\ \citenamefont
  {Schmalian}(2020)}]{AL20}%
  \BibitemOpen
  \bibfield  {author} {\bibinfo {author} {\bibfnamefont {A.}~\bibnamefont
  {Levchenko}}\ and\ \bibinfo {author} {\bibfnamefont {J.}~\bibnamefont
  {Schmalian}},\ }\href@noop {} {\bibfield  {journal} {\bibinfo  {journal}
  {Ann.\ Phys.}\ }\textbf {\bibinfo {volume} {419}},\ \bibinfo {pages} {168218}
  (\bibinfo {year} {2020})}\BibitemShut {NoStop}%
\bibitem [{\citenamefont {Moll}\ \emph {et~al.}(2016)\citenamefont {Moll},
  \citenamefont {Kushwaha}, \citenamefont {Nandi}, \citenamefont {Schmidt},\
  and\ \citenamefont {Mackenzie}}]{PJWM16}%
  \BibitemOpen
  \bibfield  {author} {\bibinfo {author} {\bibfnamefont {P.~J.~W.}\
  \bibnamefont {Moll}}, \bibinfo {author} {\bibfnamefont {P.}~\bibnamefont
  {Kushwaha}}, \bibinfo {author} {\bibfnamefont {N.}~\bibnamefont {Nandi}},
  \bibinfo {author} {\bibfnamefont {B.}~\bibnamefont {Schmidt}}, \ and\
  \bibinfo {author} {\bibfnamefont {A.~P.}\ \bibnamefont {Mackenzie}},\
  }\href@noop {} {\bibfield  {journal} {\bibinfo  {journal} {Science}\ }\textbf
  {\bibinfo {volume} {351}},\ \bibinfo {pages} {1061} (\bibinfo {year}
  {2016})}\BibitemShut {NoStop}%
\bibitem [{\citenamefont {Gooth}\ \emph {et~al.}(2018)\citenamefont {Gooth},
  \citenamefont {Menges}, \citenamefont {Kumar}, \citenamefont {S{\"{u}}\ss},
  \citenamefont {Shekhar}, \citenamefont {Sun}, \citenamefont {Drechsler},
  \citenamefont {Zierold}, \citenamefont {Felser},\ and\ \citenamefont
  {Gotsmann}}]{JG18}%
  \BibitemOpen
  \bibfield  {author} {\bibinfo {author} {\bibfnamefont {J.}~\bibnamefont
  {Gooth}}, \bibinfo {author} {\bibfnamefont {F.}~\bibnamefont {Menges}},
  \bibinfo {author} {\bibfnamefont {N.}~\bibnamefont {Kumar}}, \bibinfo
  {author} {\bibfnamefont {V.}~\bibnamefont {S{\"{u}}\ss}}, \bibinfo {author}
  {\bibfnamefont {C.}~\bibnamefont {Shekhar}}, \bibinfo {author} {\bibfnamefont
  {Y.}~\bibnamefont {Sun}}, \bibinfo {author} {\bibfnamefont {U.}~\bibnamefont
  {Drechsler}}, \bibinfo {author} {\bibfnamefont {R.}~\bibnamefont {Zierold}},
  \bibinfo {author} {\bibfnamefont {C.}~\bibnamefont {Felser}}, \ and\ \bibinfo
  {author} {\bibfnamefont {B.}~\bibnamefont {Gotsmann}},\ }\href@noop {}
  {\bibfield  {journal} {\bibinfo  {journal} {Nat.\ Commun.}\ }\textbf
  {\bibinfo {volume} {9}},\ \bibinfo {pages} {4093} (\bibinfo {year}
  {2018})}\BibitemShut {NoStop}%
\bibitem [{\citenamefont {Jaoui}\ \emph {et~al.}(2018)\citenamefont {Jaoui},
  \citenamefont {Fauqu\'{e}}, \citenamefont {Rischau}, \citenamefont {Subedi},
  \citenamefont {Fu}, \citenamefont {Gooth}, \citenamefont {Kumar},
  \citenamefont {S{\"{u}}\ss}, \citenamefont {Maslov}, \citenamefont {Felser},\
  and\ \citenamefont {Behnia}}]{AJ18}%
  \BibitemOpen
  \bibfield  {author} {\bibinfo {author} {\bibfnamefont {A.}~\bibnamefont
  {Jaoui}}, \bibinfo {author} {\bibfnamefont {B.}~\bibnamefont {Fauqu\'{e}}},
  \bibinfo {author} {\bibfnamefont {C.~W.}\ \bibnamefont {Rischau}}, \bibinfo
  {author} {\bibfnamefont {A.}~\bibnamefont {Subedi}}, \bibinfo {author}
  {\bibfnamefont {C.}~\bibnamefont {Fu}}, \bibinfo {author} {\bibfnamefont
  {J.}~\bibnamefont {Gooth}}, \bibinfo {author} {\bibfnamefont
  {N.}~\bibnamefont {Kumar}}, \bibinfo {author} {\bibfnamefont
  {V.}~\bibnamefont {S{\"{u}}\ss}}, \bibinfo {author} {\bibfnamefont {D.~L.}\
  \bibnamefont {Maslov}}, \bibinfo {author} {\bibfnamefont {C.}~\bibnamefont
  {Felser}}, \ and\ \bibinfo {author} {\bibfnamefont {K.}~\bibnamefont
  {Behnia}},\ }\href@noop {} {\bibfield  {journal} {\bibinfo  {journal} {npj\
  Quant.\ Mater}\ }\textbf {\bibinfo {volume} {3}},\ \bibinfo {pages} {64}
  (\bibinfo {year} {2018})}\BibitemShut {NoStop}%
\bibitem [{\citenamefont {Vool}\ \emph {et~al.}(2021)\citenamefont {Vool},
  \citenamefont {Hamo}, \citenamefont {Varnavides}, \citenamefont {Wand},
  \citenamefont {Zhou}, \citenamefont {Kumar}, \citenamefont {Dovzhenko},
  \citenamefont {Qiu}, \citenamefont {Garcia}, \citenamefont {Pierce},
  \citenamefont {Gooth}, \citenamefont {Anikeeva}, \citenamefont {Felser},
  \citenamefont {Narang},\ and\ \citenamefont {Yacoby}}]{UV21}%
  \BibitemOpen
  \bibfield  {author} {\bibinfo {author} {\bibfnamefont {U.}~\bibnamefont
  {Vool}}, \bibinfo {author} {\bibfnamefont {A.}~\bibnamefont {Hamo}}, \bibinfo
  {author} {\bibfnamefont {G.}~\bibnamefont {Varnavides}}, \bibinfo {author}
  {\bibfnamefont {Y.}~\bibnamefont {Wand}}, \bibinfo {author} {\bibfnamefont
  {T.~X.}\ \bibnamefont {Zhou}}, \bibinfo {author} {\bibfnamefont
  {N.}~\bibnamefont {Kumar}}, \bibinfo {author} {\bibfnamefont
  {Y.}~\bibnamefont {Dovzhenko}}, \bibinfo {author} {\bibfnamefont
  {Z.}~\bibnamefont {Qiu}}, \bibinfo {author} {\bibfnamefont {C.~A.~C.}\
  \bibnamefont {Garcia}}, \bibinfo {author} {\bibfnamefont {A.~T.}\
  \bibnamefont {Pierce}}, \bibinfo {author} {\bibfnamefont {J.}~\bibnamefont
  {Gooth}}, \bibinfo {author} {\bibfnamefont {P.}~\bibnamefont {Anikeeva}},
  \bibinfo {author} {\bibfnamefont {C.}~\bibnamefont {Felser}}, \bibinfo
  {author} {\bibfnamefont {P.}~\bibnamefont {Narang}}, \ and\ \bibinfo {author}
  {\bibfnamefont {A.}~\bibnamefont {Yacoby}},\ }\href@noop {} {\bibfield
  {journal} {\bibinfo  {journal} {Nat.\ Phys.}\ }\textbf {\bibinfo {volume}
  {17}},\ \bibinfo {pages} {1216} (\bibinfo {year} {2021})}\BibitemShut
  {NoStop}%
\bibitem [{\citenamefont {Yamanaka}\ \emph {et~al.}(2022)\citenamefont
  {Yamanaka}, \citenamefont {Okazaki},\ and\ \citenamefont {Yaguchi}}]{TY22}%
  \BibitemOpen
  \bibfield  {author} {\bibinfo {author} {\bibfnamefont {T.}~\bibnamefont
  {Yamanaka}}, \bibinfo {author} {\bibfnamefont {R.}~\bibnamefont {Okazaki}}, \
  and\ \bibinfo {author} {\bibfnamefont {H.}~\bibnamefont {Yaguchi}},\
  }\href@noop {} {\bibfield  {journal} {\bibinfo  {journal} {Phys.\ Rev.\ B}\
  }\textbf {\bibinfo {volume} {105}},\ \bibinfo {pages} {184507} (\bibinfo
  {year} {2022})}\BibitemShut {NoStop}%
\bibitem [{\citenamefont {Mott}(1936)}]{NFM36}%
  \BibitemOpen
  \bibfield  {author} {\bibinfo {author} {\bibfnamefont {N.~F.}\ \bibnamefont
  {Mott}},\ }\href@noop {} {\bibfield  {journal} {\bibinfo  {journal} {Proc.\
  R.\ Soc.\ Lond.\ A}\ }\textbf {\bibinfo {volume} {153}},\ \bibinfo {pages}
  {699} (\bibinfo {year} {1936})}\BibitemShut {NoStop}%
\bibitem [{\citenamefont {Wilson}(1938)}]{AHW38}%
  \BibitemOpen
  \bibfield  {author} {\bibinfo {author} {\bibfnamefont {A.~H.}\ \bibnamefont
  {Wilson}},\ }\href@noop {} {\bibfield  {journal} {\bibinfo  {journal} {Proc.\
  R.\ Soc.\ Lond.\ A}\ }\textbf {\bibinfo {volume} {167}},\ \bibinfo {pages}
  {580} (\bibinfo {year} {1938})}\BibitemShut {NoStop}%
\bibitem [{\citenamefont {Mott}\ and\ \citenamefont {Jones}(1958)}]{NFM58}%
  \BibitemOpen
  \bibfield  {author} {\bibinfo {author} {\bibfnamefont {N.~F.}\ \bibnamefont
  {Mott}}\ and\ \bibinfo {author} {\bibfnamefont {H.}~\bibnamefont {Jones}},\
  }\href@noop {} {\emph {\bibinfo {title} {The Theory of the Properties of
  Metals and Alloys}}}\ (\bibinfo  {publisher} {Oxford University Press},\
  \bibinfo {year} {1958})\BibitemShut {NoStop}%
\bibitem [{\citenamefont {Mueller}\ \emph {et~al.}(1970)\citenamefont
  {Mueller}, \citenamefont {Freeman}, \citenamefont {Dimmock},\ and\
  \citenamefont {Furdyna}}]{FMM70}%
  \BibitemOpen
  \bibfield  {author} {\bibinfo {author} {\bibfnamefont {F.~M.}\ \bibnamefont
  {Mueller}}, \bibinfo {author} {\bibfnamefont {A.~J.}\ \bibnamefont
  {Freeman}}, \bibinfo {author} {\bibfnamefont {J.~O.}\ \bibnamefont
  {Dimmock}}, \ and\ \bibinfo {author} {\bibfnamefont {A.~M.}\ \bibnamefont
  {Furdyna}},\ }\href@noop {} {\bibfield  {journal} {\bibinfo  {journal}
  {Phys.\ Rev.\ B}\ }\textbf {\bibinfo {volume} {1}},\ \bibinfo {pages} {4617}
  (\bibinfo {year} {1970})}\BibitemShut {NoStop}%
\bibitem [{\citenamefont {Fradin}(1974)}]{FYF74}%
  \BibitemOpen
  \bibfield  {author} {\bibinfo {author} {\bibfnamefont {F.~Y.}\ \bibnamefont
  {Fradin}},\ }\href@noop {} {\bibfield  {journal} {\bibinfo  {journal} {Phys.\
  Rev.}\ }\textbf {\bibinfo {volume} {33}},\ \bibinfo {pages} {158} (\bibinfo
  {year} {1974})}\BibitemShut {NoStop}%
\bibitem [{\citenamefont {Cutler}\ and\ \citenamefont {Mott}(1969)}]{MC69}%
  \BibitemOpen
  \bibfield  {author} {\bibinfo {author} {\bibfnamefont {M.}~\bibnamefont
  {Cutler}}\ and\ \bibinfo {author} {\bibfnamefont {N.~F.}\ \bibnamefont
  {Mott}},\ }\href@noop {} {\bibfield  {journal} {\bibinfo  {journal} {Phys.\
  Rev.}\ }\textbf {\bibinfo {volume} {181}},\ \bibinfo {pages} {1336} (\bibinfo
  {year} {1969})}\BibitemShut {NoStop}%
\bibitem [{\citenamefont {Takahashi}\ \emph {et~al.}(2016)\citenamefont
  {Takahashi}, \citenamefont {Okazaki}, \citenamefont {Ishiwata}, \citenamefont
  {Taniguchi}, \citenamefont {Okutani}, \citenamefont {Hagiwara},\ and\
  \citenamefont {Terasaki}}]{HT16}%
  \BibitemOpen
  \bibfield  {author} {\bibinfo {author} {\bibfnamefont {H.}~\bibnamefont
  {Takahashi}}, \bibinfo {author} {\bibfnamefont {R.}~\bibnamefont {Okazaki}},
  \bibinfo {author} {\bibfnamefont {S.}~\bibnamefont {Ishiwata}}, \bibinfo
  {author} {\bibfnamefont {H.}~\bibnamefont {Taniguchi}}, \bibinfo {author}
  {\bibfnamefont {A.}~\bibnamefont {Okutani}}, \bibinfo {author} {\bibfnamefont
  {M.}~\bibnamefont {Hagiwara}}, \ and\ \bibinfo {author} {\bibfnamefont
  {I.}~\bibnamefont {Terasaki}},\ }\href@noop {} {\bibfield  {journal}
  {\bibinfo  {journal} {Nat.\ Commun.}\ }\textbf {\bibinfo {volume} {7}},\
  \bibinfo {pages} {12732} (\bibinfo {year} {2016})}\BibitemShut {NoStop}%
\bibitem [{\citenamefont {Shoenberg}(1984)}]{DS84}%
  \BibitemOpen
  \bibfield  {author} {\bibinfo {author} {\bibfnamefont {D.}~\bibnamefont
  {Shoenberg}},\ }\href@noop {} {\emph {\bibinfo {title} {Magnetic Oscillations
  in Metals}}}\ (\bibinfo  {publisher} {Cambridge University Press},\ \bibinfo
  {year} {1984})\ pp.\ \bibinfo {pages} {333--352}\BibitemShut {NoStop}%
\bibitem [{\citenamefont {Stewart}(1984)}]{GRS84}%
  \BibitemOpen
  \bibfield  {author} {\bibinfo {author} {\bibfnamefont {G.~R.}\ \bibnamefont
  {Stewart}},\ }\href@noop {} {\bibfield  {journal} {\bibinfo  {journal} {Rev.\
  Mod.\ Phys.}\ }\textbf {\bibinfo {volume} {56}},\ \bibinfo {pages} {755}
  (\bibinfo {year} {1984})}\BibitemShut {NoStop}%
\bibitem [{\citenamefont {Taillefer}\ and\ \citenamefont
  {Lonzarich}(1988)}]{LT88}%
  \BibitemOpen
  \bibfield  {author} {\bibinfo {author} {\bibfnamefont {L.}~\bibnamefont
  {Taillefer}}\ and\ \bibinfo {author} {\bibfnamefont {G.~G.}\ \bibnamefont
  {Lonzarich}},\ }\href@noop {} {\bibfield  {journal} {\bibinfo  {journal}
  {Phys.\ Rev.\ Lett.}\ }\textbf {\bibinfo {volume} {60}},\ \bibinfo {pages}
  {1570} (\bibinfo {year} {1988})}\BibitemShut {NoStop}%
\bibitem [{\citenamefont {Wada}\ \emph {et~al.}(1987)\citenamefont {Wada},
  \citenamefont {Nakamura}, \citenamefont {Fukami}, \citenamefont {Yoshimura},
  \citenamefont {Shiga},\ and\ \citenamefont {Nakamura}}]{HW87}%
  \BibitemOpen
  \bibfield  {author} {\bibinfo {author} {\bibfnamefont {H.}~\bibnamefont
  {Wada}}, \bibinfo {author} {\bibfnamefont {H.}~\bibnamefont {Nakamura}},
  \bibinfo {author} {\bibfnamefont {E.}~\bibnamefont {Fukami}}, \bibinfo
  {author} {\bibfnamefont {K.}~\bibnamefont {Yoshimura}}, \bibinfo {author}
  {\bibfnamefont {M.}~\bibnamefont {Shiga}}, \ and\ \bibinfo {author}
  {\bibfnamefont {Y.}~\bibnamefont {Nakamura}},\ }\href@noop {} {\bibfield
  {journal} {\bibinfo  {journal} {J.\ Magn.\ Magn.\ Mater.}\ }\textbf {\bibinfo
  {volume} {70}},\ \bibinfo {pages} {17} (\bibinfo {year} {1987})}\BibitemShut
  {NoStop}%
\bibitem [{\citenamefont {Kondo}\ \emph {et~al.}(1997)\citenamefont {Kondo},
  \citenamefont {Johnston}, \citenamefont {Swenson}, \citenamefont {Borsa},
  \citenamefont {Mahajan}, \citenamefont {Miller}, \citenamefont {Gu},
  \citenamefont {Goldman}, \citenamefont {Maple}, \citenamefont {Gajewski},
  \citenamefont {Freeman}, \citenamefont {Dilley}, \citenamefont {Dickey},
  \citenamefont {Merrin}, \citenamefont {Kojima}, \citenamefont {Luke},
  \citenamefont {Uemura}, \citenamefont {Chmaissem},\ and\ \citenamefont
  {Jorgensen}}]{SK97}%
  \BibitemOpen
  \bibfield  {author} {\bibinfo {author} {\bibfnamefont {S.}~\bibnamefont
  {Kondo}}, \bibinfo {author} {\bibfnamefont {D.~C.}\ \bibnamefont {Johnston}},
  \bibinfo {author} {\bibfnamefont {C.~A.}\ \bibnamefont {Swenson}}, \bibinfo
  {author} {\bibfnamefont {F.}~\bibnamefont {Borsa}}, \bibinfo {author}
  {\bibfnamefont {A.~V.}\ \bibnamefont {Mahajan}}, \bibinfo {author}
  {\bibfnamefont {L.~L.}\ \bibnamefont {Miller}}, \bibinfo {author}
  {\bibfnamefont {T.}~\bibnamefont {Gu}}, \bibinfo {author} {\bibfnamefont
  {A.~I.}\ \bibnamefont {Goldman}}, \bibinfo {author} {\bibfnamefont {M.~B.}\
  \bibnamefont {Maple}}, \bibinfo {author} {\bibfnamefont {D.~A.}\ \bibnamefont
  {Gajewski}}, \bibinfo {author} {\bibfnamefont {E.~J.}\ \bibnamefont
  {Freeman}}, \bibinfo {author} {\bibfnamefont {N.~R.}\ \bibnamefont {Dilley}},
  \bibinfo {author} {\bibfnamefont {R.~P.}\ \bibnamefont {Dickey}}, \bibinfo
  {author} {\bibfnamefont {J.}~\bibnamefont {Merrin}}, \bibinfo {author}
  {\bibfnamefont {K.}~\bibnamefont {Kojima}}, \bibinfo {author} {\bibfnamefont
  {G.~M.}\ \bibnamefont {Luke}}, \bibinfo {author} {\bibfnamefont {Y.~J.}\
  \bibnamefont {Uemura}}, \bibinfo {author} {\bibfnamefont {O.}~\bibnamefont
  {Chmaissem}}, \ and\ \bibinfo {author} {\bibfnamefont {J.~D.}\ \bibnamefont
  {Jorgensen}},\ }\href@noop {} {\bibfield  {journal} {\bibinfo  {journal}
  {Phys.\ Rev.\ Lett.}\ }\textbf {\bibinfo {volume} {78}},\ \bibinfo {pages}
  {3729} (\bibinfo {year} {1997})}\BibitemShut {NoStop}%
\bibitem [{\citenamefont {Urano}\ \emph {et~al.}(2000)\citenamefont {Urano},
  \citenamefont {Nohara}, \citenamefont {Kondo}, \citenamefont {Sakai},
  \citenamefont {Takagi}, \citenamefont {Shiraki},\ and\ \citenamefont
  {Okubo}}]{CU00}%
  \BibitemOpen
  \bibfield  {author} {\bibinfo {author} {\bibfnamefont {C.}~\bibnamefont
  {Urano}}, \bibinfo {author} {\bibfnamefont {M.}~\bibnamefont {Nohara}},
  \bibinfo {author} {\bibfnamefont {S.}~\bibnamefont {Kondo}}, \bibinfo
  {author} {\bibfnamefont {F.}~\bibnamefont {Sakai}}, \bibinfo {author}
  {\bibfnamefont {H.}~\bibnamefont {Takagi}}, \bibinfo {author} {\bibfnamefont
  {T.}~\bibnamefont {Shiraki}}, \ and\ \bibinfo {author} {\bibfnamefont
  {T.}~\bibnamefont {Okubo}},\ }\href@noop {} {\bibfield  {journal} {\bibinfo
  {journal} {Phys.\ Rev.\ Lett.}\ }\textbf {\bibinfo {volume} {85}},\ \bibinfo
  {pages} {1052} (\bibinfo {year} {2000})}\BibitemShut {NoStop}%
\bibitem [{\citenamefont {Gomersall}\ and\ \citenamefont
  {Gyorffy}(1974)}]{IRG74}%
  \BibitemOpen
  \bibfield  {author} {\bibinfo {author} {\bibfnamefont {I.~R.}\ \bibnamefont
  {Gomersall}}\ and\ \bibinfo {author} {\bibfnamefont {B.~L.}\ \bibnamefont
  {Gyorffy}},\ }\href@noop {} {\bibfield  {journal} {\bibinfo  {journal} {J.\
  Phys.\ F:\ Met.\ Phys.}\ }\textbf {\bibinfo {volume} {4}},\ \bibinfo {pages}
  {1204} (\bibinfo {year} {1974})}\BibitemShut {NoStop}%
\bibitem [{\citenamefont {Miyake}\ \emph {et~al.}(1989)\citenamefont {Miyake},
  \citenamefont {Matsuura},\ and\ \citenamefont {Varma}}]{KM89}%
  \BibitemOpen
  \bibfield  {author} {\bibinfo {author} {\bibfnamefont {K.}~\bibnamefont
  {Miyake}}, \bibinfo {author} {\bibfnamefont {T.}~\bibnamefont {Matsuura}}, \
  and\ \bibinfo {author} {\bibfnamefont {C.~M.}\ \bibnamefont {Varma}},\
  }\href@noop {} {\bibfield  {journal} {\bibinfo  {journal} {Solid\ State\
  Commun.}\ }\textbf {\bibinfo {volume} {71}},\ \bibinfo {pages} {1149}
  (\bibinfo {year} {1989})}\BibitemShut {NoStop}%
\bibitem [{\citenamefont {Mineev}\ and\ \citenamefont
  {Samokhin}(2005)}]{VPM05}%
  \BibitemOpen
  \bibfield  {author} {\bibinfo {author} {\bibfnamefont {V.~P.}\ \bibnamefont
  {Mineev}}\ and\ \bibinfo {author} {\bibfnamefont {K.~V.}\ \bibnamefont
  {Samokhin}},\ }\href@noop {} {\bibfield  {journal} {\bibinfo  {journal}
  {Phys.\ Rev.\ B}\ }\textbf {\bibinfo {volume} {72}},\ \bibinfo {pages}
  {212504} (\bibinfo {year} {2005})}\BibitemShut {NoStop}%
\bibitem [{\citenamefont {Terashima}\ \emph {et~al.}(2008)\citenamefont
  {Terashima}, \citenamefont {Kimata}, \citenamefont {Uji}, \citenamefont
  {Sugawara}, \citenamefont {Kimura}, \citenamefont {Aoki},\ and\ \citenamefont
  {Harima}}]{TT08}%
  \BibitemOpen
  \bibfield  {author} {\bibinfo {author} {\bibfnamefont {T.}~\bibnamefont
  {Terashima}}, \bibinfo {author} {\bibfnamefont {M.}~\bibnamefont {Kimata}},
  \bibinfo {author} {\bibfnamefont {S.}~\bibnamefont {Uji}}, \bibinfo {author}
  {\bibfnamefont {T.}~\bibnamefont {Sugawara}}, \bibinfo {author}
  {\bibfnamefont {N.}~\bibnamefont {Kimura}}, \bibinfo {author} {\bibfnamefont
  {H.}~\bibnamefont {Aoki}}, \ and\ \bibinfo {author} {\bibfnamefont
  {H.}~\bibnamefont {Harima}},\ }\href@noop {} {\bibfield  {journal} {\bibinfo
  {journal} {Phys.\ Rev.\ B}\ }\textbf {\bibinfo {volume} {78}},\ \bibinfo
  {pages} {205107} (\bibinfo {year} {2008})}\BibitemShut {NoStop}%
\bibitem [{\citenamefont {Maurya}\ \emph {et~al.}(2018)\citenamefont {Maurya},
  \citenamefont {Harima}, \citenamefont {Nakamura}, \citenamefont {Shimizu},
  \citenamefont {Homma}, \citenamefont {Li}, \citenamefont {Honda},
  \citenamefont {Sato},\ and\ \citenamefont {Aoki}}]{AM18}%
  \BibitemOpen
  \bibfield  {author} {\bibinfo {author} {\bibfnamefont {A.}~\bibnamefont
  {Maurya}}, \bibinfo {author} {\bibfnamefont {H.}~\bibnamefont {Harima}},
  \bibinfo {author} {\bibfnamefont {A.}~\bibnamefont {Nakamura}}, \bibinfo
  {author} {\bibfnamefont {Y.}~\bibnamefont {Shimizu}}, \bibinfo {author}
  {\bibfnamefont {Y.}~\bibnamefont {Homma}}, \bibinfo {author} {\bibfnamefont
  {D.~X.}\ \bibnamefont {Li}}, \bibinfo {author} {\bibfnamefont
  {F.}~\bibnamefont {Honda}}, \bibinfo {author} {\bibfnamefont {Y.~J.}\
  \bibnamefont {Sato}}, \ and\ \bibinfo {author} {\bibfnamefont
  {D.}~\bibnamefont {Aoki}},\ }\href@noop {} {\bibfield  {journal} {\bibinfo
  {journal} {J.\ Phys.\ Soc.\ Jpn.}\ }\textbf {\bibinfo {volume} {87}},\
  \bibinfo {pages} {044703} (\bibinfo {year} {2018})}\BibitemShut {NoStop}%
\bibitem [{\citenamefont {Sato}\ \emph {et~al.}(2020)\citenamefont {Sato},
  \citenamefont {Harima}, \citenamefont {Nakamura}, \citenamefont {Maurya},
  \citenamefont {Shimizu}, \citenamefont {Homma}, \citenamefont {Li},
  \citenamefont {Honda},\ and\ \citenamefont {Aoki}}]{YJS20}%
  \BibitemOpen
  \bibfield  {author} {\bibinfo {author} {\bibfnamefont {Y.~J.}\ \bibnamefont
  {Sato}}, \bibinfo {author} {\bibfnamefont {H.}~\bibnamefont {Harima}},
  \bibinfo {author} {\bibfnamefont {A.}~\bibnamefont {Nakamura}}, \bibinfo
  {author} {\bibfnamefont {A.}~\bibnamefont {Maurya}}, \bibinfo {author}
  {\bibfnamefont {Y.}~\bibnamefont {Shimizu}}, \bibinfo {author} {\bibfnamefont
  {Y.}~\bibnamefont {Homma}}, \bibinfo {author} {\bibfnamefont {D.~X.}\
  \bibnamefont {Li}}, \bibinfo {author} {\bibfnamefont {F.}~\bibnamefont
  {Honda}}, \ and\ \bibinfo {author} {\bibfnamefont {D.}~\bibnamefont {Aoki}},\
  }\href@noop {} {\bibfield  {journal} {\bibinfo  {journal} {Phys.\ Rev.\ B}\
  }\textbf {\bibinfo {volume} {102}},\ \bibinfo {pages} {125114} (\bibinfo
  {year} {2020})}\BibitemShut {NoStop}%
\bibitem [{\citenamefont {\={O}nuki}\ \emph {et~al.}(2014)\citenamefont
  {\={O}nuki}, \citenamefont {Nakamura}, \citenamefont {Uejo}, \citenamefont
  {Teruya}, \citenamefont {Hedo}, \citenamefont {Nakama}, \citenamefont
  {Honda},\ and\ \citenamefont {Harima}}]{YO14}%
  \BibitemOpen
  \bibfield  {author} {\bibinfo {author} {\bibfnamefont {Y.}~\bibnamefont
  {\={O}nuki}}, \bibinfo {author} {\bibfnamefont {A.}~\bibnamefont {Nakamura}},
  \bibinfo {author} {\bibfnamefont {T.}~\bibnamefont {Uejo}}, \bibinfo {author}
  {\bibfnamefont {A.}~\bibnamefont {Teruya}}, \bibinfo {author} {\bibfnamefont
  {M.}~\bibnamefont {Hedo}}, \bibinfo {author} {\bibfnamefont {T.}~\bibnamefont
  {Nakama}}, \bibinfo {author} {\bibfnamefont {F.}~\bibnamefont {Honda}}, \
  and\ \bibinfo {author} {\bibfnamefont {H.}~\bibnamefont {Harima}},\
  }\href@noop {} {\bibfield  {journal} {\bibinfo  {journal} {J.\ Phys.\ Soc.\
  Jpn.}\ }\textbf {\bibinfo {volume} {83}},\ \bibinfo {pages} {061018}
  (\bibinfo {year} {2014})}\BibitemShut {NoStop}%
\bibitem [{\citenamefont {Kawai}\ \emph {et~al.}(2008)\citenamefont {Kawai},
  \citenamefont {Muranaka}, \citenamefont {Endo}, \citenamefont {Dung},
  \citenamefont {Doi}, \citenamefont {Ikeda}, \citenamefont {Matsuda},
  \citenamefont {Haga}, \citenamefont {Harima}, \citenamefont {Settai},\ and\
  \citenamefont {\={O}nuki}}]{TK08}%
  \BibitemOpen
  \bibfield  {author} {\bibinfo {author} {\bibfnamefont {T.}~\bibnamefont
  {Kawai}}, \bibinfo {author} {\bibfnamefont {H.}~\bibnamefont {Muranaka}},
  \bibinfo {author} {\bibfnamefont {T.}~\bibnamefont {Endo}}, \bibinfo {author}
  {\bibfnamefont {N.~D.}\ \bibnamefont {Dung}}, \bibinfo {author}
  {\bibfnamefont {Y.}~\bibnamefont {Doi}}, \bibinfo {author} {\bibfnamefont
  {S.}~\bibnamefont {Ikeda}}, \bibinfo {author} {\bibfnamefont {T.~D.}\
  \bibnamefont {Matsuda}}, \bibinfo {author} {\bibfnamefont {Y.}~\bibnamefont
  {Haga}}, \bibinfo {author} {\bibfnamefont {H.}~\bibnamefont {Harima}},
  \bibinfo {author} {\bibfnamefont {R.}~\bibnamefont {Settai}}, \ and\ \bibinfo
  {author} {\bibfnamefont {Y.}~\bibnamefont {\={O}nuki}},\ }\href@noop {}
  {\bibfield  {journal} {\bibinfo  {journal} {J.\ Phys.\ Soc.\ Jpn.}\ }\textbf
  {\bibinfo {volume} {77}},\ \bibinfo {pages} {064717} (\bibinfo {year}
  {2008})}\BibitemShut {NoStop}%
\end{thebibliography}%
\end{document}